\documentclass[aps,prd,preprintnumbers,groupedaddress,nofootinbib,amssymb,eqsecnum,notitlepage]{revtex4-2}
\usepackage[utf8]{inputenc}
\usepackage{graphicx}
\usepackage{dcolumn}
\usepackage{bm}
\usepackage{graphicx}
\usepackage{dcolumn}
\usepackage{amsmath}
\usepackage{amssymb}
\usepackage{comment}
\usepackage{epsf}

\usepackage{multirow}
\usepackage{epsf}
\usepackage{psfrag}
\usepackage{epsfig}
\usepackage{epstopdf}
\usepackage{threeparttable}
\usepackage{natbib}
\usepackage{bm}
\usepackage{cancel}
\usepackage[usenames,dvipsnames]{color}
\usepackage[colorlinks=true, linkcolor=BrickRed, citecolor=Blue, urlcolor=Blue, filecolor=Blue]{hyperref}
\usepackage{booktabs}
\usepackage{hhline}
\usepackage{empheq}
\usepackage{enumerate}
\usepackage{tikz}
\usetikzlibrary{arrows.meta,automata,positioning,shadows}

\def\B{\mathrm{B}}
\def\F{\mathrm{F}}

\begin{document}

\begingroup
\renewcommand\thefootnote{a}
\footnotetext{Y.K. and Z.Z. contributed equally to this work.}
\endgroup

\title{Tests of equivalence principle violation in doubly coupled dark energy }
\author{Yannick Kegel$^{1,a}$, Ziyang Zheng$^{1,a}$, Luca Amendola$^{1,2}$}
\email{yannick.kegel@stud.uni-heidelberg.de}\email{Zheng@thphys.uni-heidelberg.de}\email{l.amendola@thphys.uni-heidelberg.de}
\affiliation{$^1$ Institut f\"{u}r Theoretische Physik, Ruprecht-Karls-Universit\"{a}t Heidelberg, Philosophenweg 16, D-69120 Heidelberg, Germany}
\affiliation{$^2$ The Center for Astrophysics and Space Science (CASS), NYUAD Research Institute, New York University Abu Dhabi, PO Box 129188, Abu Dhabi, UAE}  
\begin{abstract}
Recent work has shown that a simple form of coupled dark energy  with a constant coupling produces a good fit to the DESI data. In this scenario, dark matter, but not baryons, couples to a scalar field, thereby violating the weak equivalence principle (EP). We first generalize the coupled dark energy model by adding a nonminimal coupling between the scalar field and gravity, and dub the resulting model “doubly coupled dark energy”. We then construct a parameter $E_P$ in this model that quantifies EP violation on cosmological scales and can be measured with large galaxy surveys in a model-independent way. Finally, we forecast constraints on the couplings  for present and future spectroscopic galaxy surveys, namely DESI, Euclid, and SKA Phase 2 (SKA2). To keep the analysis as general as possible, we leave the initial power spectrum shape and the galaxy bias free to vary in several bins. Among these surveys, SKA2 could provide the tightest constraint on the coupling strength in the coupled dark energy scenario, with a relative \(1\sigma\) uncertainty of \(22\%\). The constraints weaken when both couplings are varied, but can be further improved by increasing the number of galaxies detected.
\end{abstract}
\maketitle

\section{Introduction}

The recent DESI data analysis \cite{DESI:2024mwx, DESI:2025zgx} provided tantalizing hints for dynamical dark energy, i.e. dark energy with a variable equation of state (EoS). The equation of state alone, however, does not capture the entire phenomenology of dark energy, and models based on scalar fields, somewhat analogous to inflationary scalar fields, have been proposed and examined  \cite{Ramadan:2024kmn, Wolf:2024eph, Akrami:2025zlb, Cline:2025sbt, DESI:2025fii, Bhattacharya:2024hep, Bayat:2025xfr, Berghaus:2024kra, Bhattacharya:2024kxp, Urena-Lopez:2025rad, Gialamas:2025pwv, Alestas:2024eic}. In order to mimic the DESI best fit evolving EoS $w(z)$, however, these models need to reproduce the so-called ``phantom crossing'', i.e. a transition from $w<-1$ to $w>-1$ at some recent epoch. This behavior cannot be obtained in ordinary scalar field models without at the same time creating instabilities \cite{Vikman:2004dc}. 
Interestingly, non-minimally coupled scalar field models that introduce an interaction of the scalar field with matter seem to provide a good fit to current cosmological data while naturally producing an effective phantom crossing without introducing ghost or gradient instabilities~\cite{Wolf:2025jed, Ye:2024ywg}. Such models, however, automatically generate a fifth force acting on baryonic matter, that is tightly constrained by laboratory experiments and Solar System tests \cite{Will:2014kxa, Bertotti:2003rm} (except if new ingredients are added to screen the fifth force on small scales \cite{Khoury:2003rn, Nicolis:2008in, Hinterbichler:2010es}). An elegant solution to evade those constraints while still solving the ``phantom crossing'' problem, is to couple only dark matter to the scalar field, as shown in~\cite{Das:2005yj, Gomez-Valent:2026ept, Chakraborty:2024xas, Chakraborty:2025syu, Antusch:2026ldp}.

Perhaps the simplest and one of the most widely studied models of this type is a dark matter-dark energy interaction, namely the coupled dark energy model. In its basic incarnation, this model depends on a single constant coupling $Q_{\rm dm}$ and on a potential, for instance a power law or an exponential. At the perturbation level, this model introduces, on sub-horizon scales, a Yukawa correction to the effective gravitational potential \cite{Amendola:2003wa, Amendola:2019laa}, which for a point source of mass $M$ has the following form in real space
\begin{equation}
    \hat\Psi(r)=-\frac{GM}{r}(1+2Q_{\rm dm}^2 e^{-mr})\label{eq:ny} \,.
\end{equation}
Here, $M$ and the scalar field mass $m$ generally depend on the scalar field background value $\phi$, which, in turn, obeys a Klein-Gordon equation.  The value of $Q_{\rm dm}$ has been constrained in the literature several times, and a peak away from zero in its marginalized posterior was already found in some analyses before DESI~\cite{Pettorino_2012, Pettorino:2013oxa, Gomez-Valent:2020mqn, Xia:2013nua}. Analyses including DESI data have also found hints of a nonzero coupling, although their significance depends on the datasets and assumptions adopted~\cite{Chakraborty:2024xas, Gomez-Valent:2026ept, Wang:2025znm, Li:2026xaz, Babu:2026tfr, Ganguly:2026mgu}. 
If this evidence is confirmed by additional data, e.g.~Euclid, the dark-dark interaction could become a new standard cosmological model.

The coupled dark energy model contains a startling prediction: that the weak equivalence principle is  broken at cosmological scales. That is, baryons and dark matter fall differently in a gravitational potential: while baryons remain uncoupled and feel standard  gravity, following the geodesics of the metric, dark matter feels an extra interaction mediated by the scalar field and therefore deviates from the geodesics of the metric $g_{\mu\nu}$. Among the various effects induced by this extra interaction, the dark matter particle mass depends on the field  $\phi$ and therefore varies with time, cf. Refs \cite{Gomez-Valent:2022bku, Gomez-Valent:2026ept} (and in principle also with space, but this would occur only beyond linear theory).

Recently, a direct way to test the EP in large scale structure has been identified \cite{Bonvin:2018ckp, Bonvin:2020cxp, EP, Zheng:2025oiq}. The main idea is to employ the gravitational redshift effect to measure the gravitational potential and compare this with the effective potential $\hat\Psi$ mentioned above. This requires carefully incorporating horizon-scale relativistic corrections in the galaxy number count fluctuations \cite{Yoo:2009au, Bonvin:2011bg, Challinor:2011bk, Jeong:2011as}, which can be extracted from the two-point correlation function or the power spectrum by cross-correlating two (or more) sub-populations of galaxies with different biases \cite{McDonald:2009ud, Yoo:2012se, Croft:2013taa, Bonvin:2013ogt, Bonvin:2014owa, Gaztanaga:2015jrs}. This technique, multi-tracer with relativistic corrections, has been exploited in Ref.~\cite{EP} to provide the first forecasts of EP-violation in present and future surveys, in a way that is independent of (i.e., marginalized over) the power spectrum shape, the background cosmology, and the perturbation growth rate.

This paper is devoted to forecasting the performance of this cosmological EP test in the context of coupled dark energy (CDE) and its generalization to doubly coupled dark energy (DCDE). In the latter model, the scalar field is conformally coupled to dark matter and nonminimally coupled to gravity. We consider the observable $E_P$ identified in Ref.~\cite{EP}, which provides a null test of the EP and can be measured in a model-independent way from galaxy power spectra. We express this quantity in the CDE and DCDE models, obtaining a general form that helps understand the roles of the various sectors of the theory's Lagrangian. The model reduces to CDE when the nonminimal gravitational coupling is switched off. We then perform Fisher forecasts on these couplings for DESI, Euclid and SKA2 through a projection from $E_P$ and additional background quantities. Our constraints on these couplings are  model-independent in the sense that the underlying analysis leaves the shape of the power spectrum free to vary in several $k$-bins, and the galaxy bias also free to vary in redshift bins.
This preserves the robustness of the analysis and allows us to estimate how well present and future surveys can constrain these couplings and $E_P$ with galaxy multi-tracing and relativistic corrections, and hence assess the amount of EP-violation in our Universe.

\section{Evolution equations in a doubly coupled dark energy model} \label{sec: model}

In this section, we derive the evolution equations for the DCDE model, where dark matter and the metric are conformally coupled to a scalar field $\phi$. As mentioned above, a violation of the weak equivalence principle is already present when the scalar field couples only to dark matter, and it does not require the second coupling to the metric. However, the advantage of studying this more general doubly coupled model is that it provides insight about the impact of each coupling on cosmological observables and on their interplay regarding our test of the equivalence principle. In practice, the double coupling is introduced via two conformal factors $f_1$ and $f_2$. To construct the simplest model of this kind with constant couplings, we choose appropriate functions $f_1,f_2$. More general field-dependent couplings are of course possible, cf. Refs. \cite{Amendola:2003wa, Baldi:2010vv}. The corresponding action of a scalar-tensor theory, where a dynamical scalar field \(\phi\) is nonminimally coupled to both gravity (cf. Ref. \cite{Amendola:1999qq}) and the dark matter sector (cf. Ref. \cite{Amendola:1999er}), is given by:
\begin{align} \label{eq: action_full}
    S &= \int d^4x \sqrt{-g} \left[ \frac{1}{2 \kappa} f_1(\phi)R+L_{\phi}(g_{\mu\nu})+ L_{\rm b}(g_{\mu\nu}, \psi_{\rm b})+L_{\rm dm,}\left(f_2(\phi)g_{\mu\nu}, \psi_{\rm DM}\right)\right] \,   ,
\end{align}
where $\psi_{\rm b}$ and $\psi_{\rm dm}$ are the fields of baryon and dark matter particles considered to be fermions, $\kappa=8\pi G$ is the inverse of the squared, reduced Planck mass, and
\begin{align}
    L_{\phi}(g_{\mu\nu})=-\frac{1}{2}(1-6  Q^2)f_1(\phi)\nabla_\mu \phi\nabla^\mu\phi-V(\phi)\,,
\label{eq:Q}\end{align}
where $ Q$ is a constant. The form of Eq.~(\ref{eq:Q}) is chosen so that 
$Q$ will appear as the strength of the fifth-force induced by the coupling $f_1$, as will be shown later.
Specifically, $f_1$ induces a nonminimal coupling to the Ricci scalar, thereby modifying the gravitational sector, while $f_2$ introduces a direct coupling to dark matter through a conformal rescaling of the metric. While $  Q$ represents a universal modification of gravity, the equivalence principle is explicitly violated only when $f_2 \neq \mathrm{const.}$\footnote{An additional coupling to the baryons is of course also possible, which would modify the baryon masses or, more in general, cause baryons to deviate from the geodesics of the metric $g_{\mu\nu}$. In this scenario the cosmological observations would not be directly comparable to the theory predictions. In any case, after a conformal transformation, the baryon coupling can be absorbed into the other couplings $f_1$ and $f_2$. }

Varying the action \eqref{eq: action_full} with respect to the metric $g_{\mu\nu}$ and the scalar field $\phi$, one obtains the modified Einstein field equations and the modified Klein-Gordon equation:
\begin{equation}
     f_1G_{\mu\nu}+g_{\mu \nu} \Box f_1-\nabla_{ \mu} \nabla_{\nu} f_1 = \kappa  (T_{\mu \nu}^{\rm dm}+T_{\mu \nu}^{\rm b}+T_{\mu \nu}^{\rm \phi}) \,,
    \label{eq:back2} 
\end{equation}
\begin{equation}
    \Box\phi+\frac{1}{2}\frac{f_{1,\phi}}{f_1}\nabla_{\mu}\phi\nabla^\mu\phi
     = \frac{1}{(1-6Q^2)f_1}\left(V_{, \phi}-\frac{1}{2\kappa}f_{1,\phi}R -\frac{1}{2}\frac{f_{2,\phi}}{f_2}T^{\rm dm}\right) \,, \label{eq:back1}  
\end{equation}
where $G_{\mu\nu}=R_{\mu\nu}-\frac{1}{2}Rg_{\mu\nu}$; $\Box\phi\equiv\nabla_\mu \nabla^\mu\phi$ is the the d'Alembertian operator acting on \(\phi\); the superscripts $\{\rm b,dm\}$ denote the baryons and the dark matter components, respectively; $T^{\rm dm}$ is the trace of the dark matter energy-momentum tensor; and $f_{i,\phi}=df_i/d\phi; V_{,\phi} = dV/d\phi$.

\begin{figure}[]
\centering
\begin{tikzpicture}[scale=1.4, transform shape, 
  shorten >=1pt,node distance=2cm,on grid,>={Stealth[round]},thick,
  every state/.style={
  draw=none,
  text=black,
  circular drop shadow,
  minimum size=28pt,
  inner sep=0pt
},
  initial/.style={fill=cyan!20},
  accepting/.style={fill=red!20},        
  dmstyle/.style={fill=purple!13!blue!13},
  dmstyle2/.style={fill=yellow!30!red!25}]]

\node[state,initial]  (q_0)                {$R$};
\node[state,accepting](q_2) [left=of q_0]  {$\phi$};    
\node[state,dmstyle]  (q_3) [below=of q_0]  {$L_{\rm b}$}; 
\node[state,dmstyle2]  (q_1) [left=of q_3]  {$L_{\rm dm}$}; 

    \path[<->] 
    (q_0) edge node [above]      {$ Q(\phi)$}         (q_2);
    \path[<->] 
    (q_1) edge node [left]      {$Q_{\rm dm}(\phi)$}         (q_2);
\path[<->, dashed, draw=black!50] 
  (q_0) edge node[above right] {\footnotesize Minimal} (q_3);
\path[<->, dashed, draw=black!50] 
  (q_0) edge node[below right] {\footnotesize Coupling} (q_3);
    \path[<->, dashed, draw=black!50] 
  (q_1) edge (q_0);
\end{tikzpicture}
\caption{Schematic illustration of interactions between the Ricci scalar $R$, dark matter/baryon Lagrangian $L_{\rm dm}$/$L_{\rm b}$, and the scalar field $\phi$. The dashed arrow indicates minimal coupling via gravity, while the solid arrows represent direct couplings through the functions $ Q(\phi)$ and $Q_{\rm dm}(\phi)$.}
\label{fig:Coupling representation}
\end{figure}
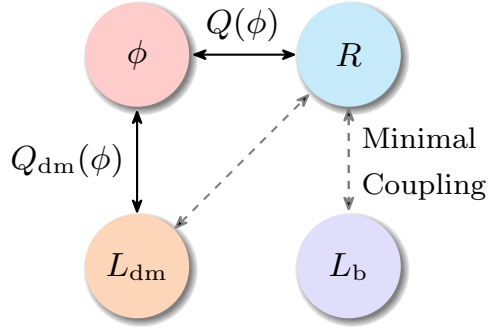
We consider a flat, linearly perturbed Friedmann-Lema\^itre-Robertson-Walker (FLRW) metric in the conformal Newtonian gauge, where the two scalar degrees of freedom are characterized by the gravitational potentials $\Psi$ and $\Phi$,
\begin{align} \label{eq: FLRW}
    ds^2=a^2\left[-(1+2 \Psi )d\eta^2+(1-2 \Phi )dx^idx_i\right] \, ,
\end{align}
where $a$ is the scale factor and $\eta$ is the conformal time. Substituting the metric Eq. \eqref{eq: FLRW} and assuming pressureless dark matter and baryons, one can obtain the Friedmann equation,
\begin{equation}
    3H^2f_1 = \kappa\left[\rho_{\rm dm}+\rho_{\rm b}+\frac{1}{2}(1-6Q^2)f_1H^2\phi'^2+V(\phi)\right] - 3H^2f_{1,\phi}\phi' \,,
\end{equation}
where a prime denotes differentiation with respect to the $e$-folding time $N = \ln a$. Note that the background energy density associated with the scalar field is given by $\rho_{\phi} = \frac{1}{2}(1-6Q^2)f_1(\phi)H^2\phi'^2+V(\phi)$, and the ``effective" dark energy density reads 
\begin{equation}
\rho_{\rm DE, eff} = \frac{1}{2}(1-6Q^2)f_1(\phi)H^2\phi'^2+V(\phi)- \frac{3H^2f_{1,\phi}\phi'}{\kappa} \,,
\end{equation}
such that
\begin{equation} \label{eq: Friedmann_effctive}
    3H^2f_1 = \kappa\left(\rho_{\rm dm}+\rho_{\rm b}+\rho_{\rm DE,eff}\right) \,,
\end{equation}
and one can define
\begin{equation}
    \tilde{\Omega}_{\rm dm}= \frac{\kappa\rho_{\rm dm}}{3H^2f_1}; \quad \tilde{\Omega}_{\rm b}= \frac{\kappa\rho_{\rm b}}{3H^2f_1}; \quad  \tilde{\Omega}_{\rm DE, eff}= \frac{\kappa\rho_{\rm DE, eff}}{3H^2f_1}  \,,
\end{equation}
to ensure that $\tilde{\Omega}_{\rm dm} + \tilde{\Omega}_{\rm b}+\tilde{\Omega}_{\rm DE, eff} =1$. See also Ref. \cite{Tsujikawa:2008uc} in which these density fractions are defined in a similar way.
The background Klein-Gordon equation is given by
\begin{equation}
\phi''+\left(3+\frac{H'}{H}\right)\phi'+\frac{1}{2}\frac{f_{1,\phi}}{f_1}\phi'^2
=-\frac{1}{(1-6Q^2)\,f_1 H^2}\left[V_{,\phi}-\frac{3f_{1,\phi}}{\kappa}H^2\left(2+\frac{H'}{H}\right)+\frac{f_{2,\phi}}{2f_2}\rho_{\rm dm}\right]  \,,
\label{eq:KG_BG}
\end{equation}
Defining $Q_{\rm dm}(\phi)\equiv\frac{f_{2,\phi}}{2f_2}$, the conservation equations for baryons, dark matter, and the effective dark energy component can also be derived from the action \eqref{eq: action_full}: 
\begin{align} 
    \nabla^{ \nu}T_{ \nu \mu}^{\rm b} &=0  \,, \label{eq: con_b}\\
    \nabla^{ \nu }T_{\nu \mu}^{\rm dm} &= Q_{\rm dm}T^{\rm dm} \nabla_{\mu} \phi \label{eq: con_dm}   \,, \\
    \nabla^{ \nu }T_{\nu \mu}^{\rm DE, eff} &= -Q_{\rm dm} T^{\rm dm} \nabla_{\mu}\phi +  \frac{1}{\kappa}G_{\nu\mu}\nabla^\nu f_1  \,.
\end{align}
The time components of the above conservation equations at the background level then lead to
\begin{align} 
    \rho_{\rm b}'+3\rho_{\rm b}&=0 \,, \label{eq: continuity_baryon} \\
    \rho_{\rm dm}'+3\rho_{\rm dm}&=Q_{\rm dm}\rho_{\rm dm}\phi' \,, \label{eq: continuity_dark_matter} \\
    \rho_{\rm DE,eff}'+3(\rho_{\rm DE,eff}+p_{\rm DE,eff})&=-Q_{\rm dm}\rho_{\rm dm}\phi'+\frac{3H^2}{\kappa}f_{1,\phi}\phi' \,. \label{eq: continuity_DE_eff}
\end{align}
Solving Eq. \eqref{eq: continuity_dark_matter}, we have
\begin{equation}
  \rho_{\rm dm}(N) = m_{\rm dm}(N)n_{\rm dm}(N) = \rho_{\rm dm,0}e^{-3N}e^{\int_{\phi_0}^{\phi}Q_{\rm dm}(\hat{\phi})d\hat{\phi}} \,,
\end{equation}
where a subscript "0" denotes the quantities at the current epoch. Assuming a conserved total number of dark matter particles, the particle mass varies due to the presence of the coupling $Q_{\rm dm}$, and evolves as $m_{\rm dm}(N)=m_{\rm dm,0}e^{\int_{\phi_0}^{\phi}Q_{\rm dm}(\hat{\phi})d\hat{\phi}}$.

At the perturbation level, the perturbed scalar field $\phi_{(\rm p)}$, the density $\rho_{\rm m, (\rm p)}$ and four-velocity $u_{\rm m,(\rm p)}^{\mu}$ of the matter contents, on the other hand, are expanded as
\begin{align}
    \phi_{(\rm p)}= \phi + \varphi  \; ; \quad \rho_{\rm m, (p)} = \rho_{\rm m}(1 + \delta_{\rm m}) ; \quad  u_{\rm m,(\rm p)}^{\mu}=\frac{1}{a}(1-\Psi, v^i_{\rm m})
\end{align}
with a subscript (p) denoting the perturbed quantities, $v^i=dx^i/d\eta$ the  peculiar velocity defined using the comoving coordinates $x^i$, and ${\rm m} = \{\rm b,dm\}$.
Inserting Eq. \eqref{eq: FLRW} into the matter conservation equations \eqref{eq: con_b} and \eqref{eq: con_dm}, the continuity and Euler equations in Fourier space \footnote{Throughout this work we employ the Fourier convention 
\begin{equation}
\tilde{f}(\textbf{k},t)=\int d^3\textbf{x}f(\textbf{x},t)e^{-i\textbf{k}\cdot\textbf{x}}; \quad f(\textbf{x},t)=\int \frac{d^3\textbf{k}}{(2\pi)^3}\tilde{f}(\textbf{k},t)e^{i\textbf{k}\cdot\textbf{x}}\nonumber  \,.
\end{equation}}, are given by 
\begin{align}
    \delta_{\rm b}' &= -\theta_{\rm b} + 3 \Phi'\label{Cont baryons} \,,\\ 
    \theta_{\rm b}' &+ \theta_{\rm b}(1+\frac{\mathcal{H}'}{\mathcal{H}}) -\frac{\Psi}{\lambda^2}=0 \label{eq:Euler_baryon} \,,\\ 
    \delta_{\rm dm}' & + \theta_{\rm dm} -3 \Phi' = Q_{\rm dm}\varphi' +Q_{\rm dm,\phi}  \phi' \varphi \label{Cont dm} \,, \\
   \theta_{\rm dm}' &+ \theta_{\rm dm}(1+\frac{\mathcal{H}'}{\mathcal{H}}+Q_{\rm dm}\phi') - \frac{Q_{\rm dm}\varphi+\Psi}{\lambda^2} = 0 \label{eq:Euler_dm}  \,,
\end{align}
where $\theta_{\rm m}$ is the rescaled velocity divergence $\theta_{\rm m}=ik_i v^i_{\rm m}/\mathcal{H}$ with wave vector $k$ and $\mathcal{H}=aH$ being the conformal Hubble function. We also introduce $\lambda =\mathcal{H}/k$ to characterize the scale dependence and define $Q_{{\rm dm},\phi}\equiv dQ_{\rm dm}/d\phi$. On the other hand, from the perturbed Einstein equations we obtain
\begin{align}
    \Phi &= -\frac{\lambda^2}{2 f_1}\Bigg[ 2\hat{V} \Psi+  \varphi \hat{V}_{,\phi}-3\varphi f_{1,\phi\phi}\phi'+3(\delta_{\rm b}\Omega_{\rm b}+\delta_{\rm dm}\Omega_{\rm dm}) +\Phi'(6 f_1+3\phi'f_{1,\phi}) \label{First Einstein} \nonumber \\
    & - 3 f_{1,\phi} (\varphi+\varphi')+6\Psi (\Omega_{\rm b}+\Omega_{\rm dm})+\kappa(1-6 Q^2)\left(f_1\phi'\varphi'+\frac{f_{1,\phi}\varphi\phi'^2}{2}\right)\Bigg]+\varphi \frac{f_{1,\phi}}{2 f_1} \,, \\ 
    \Psi &=  \Phi - \varphi \frac{f_{1,\phi}}{f_1}   \label{Second Einstein}   \,,
\end{align}
with $\hat{V}=Va^2/\mathcal{H}^2$. From now on, we adopt the reduced Planck units where $\kappa = c = 1$. Also, to avoid confusion, notice that $\Omega_{\rm m}(N) =\kappa\rho_{\rm m}/3H^2=f_1\tilde{\Omega}_{\rm dm}$ is a time-dependent quantity, not the present values. So far, we have allowed $Q_{\rm dm}$ to vary with time. Hereafter we specialize to constant couplings $Q=-\frac{f_{1,\phi}}{2f_1}$ and $Q_{\rm dm}=\frac{f_{2,\phi}}{2f_2}$, because in this case one obtains a Yukawa correction in Eq.~(\ref{eq:ny}) with a constant coupling. Then the general solutions are $f_1(\phi)=f_{1,0}e^{-2Q(\phi-\phi_0)},  f_2(\phi)=f_{2,0}e^{2Q_{\rm dm}(\phi-\phi_0)}$, with $\phi_0$ the present-day field value and $f_{i,0}\equiv f_i(\phi_0)$. We adopt the normalization $f_{1,0}=f_{2,0}=1$. In addition, we can further set $\phi_0=0$ without loss of generality. With these conventions, the coupling functions become $f_1=e^{-2Q\phi}$ and $f_2=e^{2Q_{\rm dm}\phi}$.

On sub-horizon scales ($\lambda\ll 1$), we adopt the quasi-static assumption (QSA), as decribed in Ref. \cite{Amendola2019-ey},  by implying that spatial derivatives dominate over the time derivatives of slowly evolving perturbations, i.e. $X \sim X' \sim X''$ for all perturbation quantities $X \in \{\delta, \theta, \Psi, \Phi\}$. In this case $X', X'' \sim X \ll \lambda^{-2}X$. 
\footnote{The perturbed Klein-Gordon equation generally admits rapidly oscillating homogeneous solutions, whose time derivatives cannot be neglected. However, averaging over these oscillations isolates the slowly evolving, matter-sourced component, for which the QSA is valid. In the following, we omit the averaging brackets for simplicity.}
Consequently, the perturbed Klein-Gordon equation reads,
\begin{align}
    \left(1-6 Q^2+\frac{a^2m_{\phi}^2}{k^2}\right)\varphi = -3\lambda^{2} Q_{\rm dm}\tilde{\Omega}_{\rm dm}\delta_{\rm dm}- Q\left(2\Psi - 4 \Phi\right)\, , \label{KG in between}
\end{align}
where $m_\phi^2= \left(V_{\rm ,\phi\phi}+2QV_{\rm , \phi}+2QQ_{\rm dm}\rho_{\rm dm}\right)/f_1(\phi)$. Moreover, under the QSA Eq.~\eqref{First Einstein} reduces to 
\begin{equation}
\Phi = -\frac{3\lambda^2}{2}(\delta_{\rm b}\tilde{\Omega}_{\rm b}+\delta_{\rm dm}\tilde{\Omega}_{\rm dm})-Q\varphi \,,
\end{equation}
and Eq.~\eqref{KG in between} can be brought into the convenient form,
\begin{align}
    \varphi = -3\lambda^2Y(k)\Big[(Q+Q_{\rm dm})\tilde{\Omega}_{\rm dm}\delta_{\rm dm}+ Q\tilde{\Omega}_{\rm b}\delta_{\rm b}\Big]\;; \quad Y(k)=\frac{k^2}{k^2+a^2m_{\phi}^2} \label{Poisson phi} \,.
\end{align}
Thus, Eq.~\eqref{Second Einstein} can be re-written for the scalar field $\Psi$:
\begin{align}
    \Psi = -\frac{3\lambda^2}{2}\Big[\big(2QY(k)( Q+Q_{\rm dm})+1\big)\tilde{\Omega}_{\rm dm}\delta_{\rm dm}+\big(2Q^2Y(k)+1\big)\tilde{\Omega}_{\rm b}\delta_{\rm b}\Big] \,. \label{Poisson psi}
\end{align}
These are the Poisson equations for the perturbation quantities $\varphi$ and $\Psi$.  

To determine the growth of matter perturbations at deep subhorizon scales, the continuity equations \eqref{Cont baryons} and \eqref{Cont dm} can be simplified to $\delta_{\rm b}' = -\theta_{\rm b}$ and $\delta_{\rm dm}' = -\theta_{\rm dm}$, respectively, because from the Poisson equations one sees that $\Psi,\varphi\sim \lambda^2 \delta_{\rm m}$. Taking the derivative of  \eqref{Cont baryons} and \eqref{Cont dm} and replacing $\theta_{\rm b},\theta_{\rm dm}$ and their derivatives, the growth equations for baryons and dark matter takes the following form,
\begin{align}
    \delta_{\rm b}'' &+ \delta_{\rm b}'(1+\frac{\mathcal{H}'}{\mathcal{H}}) +\frac{\Psi}{\lambda^2}=0 \,, \label{Euler b}  \\ 
   \delta_{\rm dm}'' &+ \delta_{\rm dm}'(1+\frac{\mathcal{H}'}{\mathcal{H}} +Q_{\rm dm}\phi') + \frac{\Psi + Q_{\rm dm} \varphi}{\lambda^2} = 0  \,.  \label{Euler dm old}
\end{align}
Plugging in the gravitational potentials we have
\begin{align}
    \delta_{\rm b}'' &+ (1+\frac{\mathcal{H}'}{\mathcal{H}})\delta_{\rm b}' -\frac{3}{2}\Big[\big(1+2QY( Q+Q_{\rm dm})\big)\tilde{\Omega}_{\rm dm}\delta_{\rm dm}+\big(1+  2Q^2Y\big)\tilde{\Omega}_{\rm b}\delta_{\rm b}\Big] =0 \label{Euler b_2} \,, \\ 
   \delta_{\rm dm}'' &+ (1+\frac{\mathcal{H}'}{\mathcal{H}} +Q_{\rm dm}\phi')\delta_{\rm dm}' -\frac{3}{2}\left\{ \left[1+2Y(Q+Q_{\rm dm})^2\right]\tilde{\Omega}_{\rm dm}\delta_{\rm dm}+\left[1+2QY(Q+Q_{\rm dm})\right]\tilde{\Omega}_{\rm b}\delta_{\rm b} \right\} = 0 \label{Euler dm_2} \,.
\end{align}
As expected, for $Q=0$, these equations reduce to the growth equations in the CDE scenario, see e.g., Refs.~\cite{Gomez-Valent:2022bku, Gomez-Valent:2020mqn}

The effective potential acting on dark matter test particles mentioned in the introduction is simply $\hat\Psi_{\rm dm}\equiv \Psi+Q_{\rm dm}\varphi$, while for baryon test particles it is $\hat\Psi_{\rm b}\equiv \Psi$. As anticipated, in real space the gravitational potential is modified by the term $Y(k)$ and the Newtonian potential around a point-like dark matter particle acquires a Yukawa correction $  2(  Q+Q_{\rm dm})^2e^{-m(\phi)r}$, while the correction for a baryon particle is $2 Q^2 e^{-m(\phi)r}$. We present this explicitly in App. \ref{app:proof}.  One can then think of dark matter particles as carrying a ``charge'' for the field interaction $Q+Q_{\rm dm}$, and of baryons as carrying a charge $Q$.

\section{Testing the Weak Equivalence Principle}

In this section, we derive a combination of quantities in the model described in Sec. \ref{sec: model} that can be employed to test the EP with galaxy clustering, as shown in Refs.~\cite{EP, Zheng:2025oiq}.

For a generic model, the Poisson equation for the gravitational potential $\Psi$ takes the form,
\begin{equation}
    \Psi = -\frac{3}{2} \sum_{\rm m}\mu_{G,{\rm m}}(k,z) \lambda^2 \Omega_{\rm m} \delta_{\rm m} \,, 
\end{equation}
where the summation runs over the matter species ${\rm m}=\{\rm dm,b\}$. Here, $\mu_{G,\rm m}$ is the effective gravitational ``constant'' that could possibly depend on both time and scale, with $\mu_{G,\rm m} \neq 1$ quantifying departures from general relativity. The previously derived Poisson equation \eqref{Poisson psi} can be used to identify that,  
\begin{align}
    \mu_{G,\rm dm}(k,z) &= \frac{1}{f_1}\left[Y(k)(2 Q^2+2Q_{\rm dm} Q)+1\right] \label{Mu dm}  \,,   \\
     \mu_{G,\rm b}(k,z) &= \frac{1}{f_1}\left[Y(k)2Q^2+1\right] \label{Mu b}  \,.
\end{align}
In Ref.~\cite{EP}, it was shown that a parameter, denoted as $E_P$, can be identified and measured in galaxy clustering surveys  when relativistic corrections (in particular, terms associated with the gravitational redshift) are taken into account (see details in Sec. \ref{sec: forecast}).
This parameter $E_P$ measures the deviation from the EP between baryons and dark matter, and can be estimated in a model-independent way, i.e. without choosing the initial conditions of the Universe or a particular cosmological model, either at the background or at the perturbation level. As we now show, for this DCDE model, $E_P$ can be explicitly constructed from the Poisson and the Euler equations for dark matter and baryons.

Proposed in \cite{Bonvin:2018ckp}, the most general cosmological Euler equation for dark matter in the modified gravity scenario can be expressed by introducing two phenomenological free functions (see, e.g., Refs \cite{Castello:2022uuu, EP, Zheng:2025oiq}), that become nonzero when deviating from the standard  $\Lambda$CDM scenario,
 \begin{equation}
     \theta_{\rm dm}' + \theta_{\rm dm}\left[1+\frac{\mathcal{H}'}{\mathcal{H}}+\Theta_{\rm dm}(k,z)\right] - \Big[1+\Gamma_{\rm dm}(k,z)\Big]{\lambda^{-2}}\Psi = 0 \,.
 \end{equation}
Here, the functions $\Theta_{\rm dm}, \Gamma_{\rm dm}$ correspond to an additional friction term and to the fifth force acting on dark matter, respectively. By comparing with Eq. \eqref{eq:Euler_dm} one can find that \cite{Amendola:2003wa}
\begin{align}
    \Theta_{\rm dm}=Q_{\rm dm}\phi' \; ; \quad \Gamma_{\rm dm}=\frac{Q_{\rm dm}\varphi}{\Psi} \,.\label{Theta dm}
\end{align}
As expected, $\Theta$ and $\Gamma$ both vanish when $Q_{\rm dm}\to 0$, i.e. when the EP is satisfied. Substituting $\varphi$ and $\Psi$ with Eqs. \eqref{Poisson phi} and \eqref{Poisson psi}, $\Gamma_{\rm dm}$ becomes
\begin{align}
    \Gamma_{\rm dm}=2Q_{\rm dm}Y(k)\frac{ \left(Q+Q_{\rm dm}\right)\Omega_{\rm dm}+ Q\Omega_{b}b}{\Big[2QY(k)(Q+Q_{\rm dm})+1\Big]\Omega_{\rm dm}+\Big[2Q^2Y(k)+1\Big]\Omega_{\rm b}b} \label{Gamma dm}  \,,
\end{align}
with the time-dependent bias function $b(z)$ defined by $\delta_{\rm b}=b(z)\delta_{\rm dm}$. 
On linear scales, it can be obtained by integrating Eqs.~\eqref{Euler b} and \eqref{Euler dm old}. Notice that this bias is exclusively due to the EP violation and does not include  fluidodynamical effects that are expected to be important at non-linear scales.
Referring only to the dark matter component, $E_{P}$ is defined in \cite{EP} in the following way,
\begin{align}
    E_{P} = 1 + \Theta_{\rm dm} - \frac{3 \Omega_{\rm dm} \mu_{G,\rm dm} \Gamma_{\rm dm}}{2f_{\rm g}} \,,
\end{align}
where $f_g\equiv \delta'_{\rm dm}/\delta_{\rm dm}$ is the dark matter growth rate. 
It is important to remark that neither $\Theta_{\rm dm}$ nor $\Gamma_{\rm dm}$ are directly observable, unless one first assume a specific cosmology, while $E_P$ is a directly observable quantity, as will be shown in Sect. \ref{sec: forecast}.
Whenever $E_{P}\not= 1$, the EP is violated, whether originating from non-standard gravitational interactions or non-minimal couplings in the dark sector. 
Combining Eqs.~\eqref{Mu dm}, \eqref{Theta dm} and \eqref{Gamma dm}, $E_P$ for dark matter can be expressed in the most general form,
\begin{align}
    E_{P}=1 + Q_{\rm dm}\phi'-\frac{3\Omega_{\rm dm}}{2f_gf_1}\frac{Q_{\rm dm}Y(k)\Big[2(Q+Q_{\rm dm})\Omega_{\rm dm}+2 Q\Omega_{\rm b}b\Big]\Big[2QY(k)(Q+Q_{\rm dm})+1\Big]}{\Big[2QY(k)(Q+Q_{\rm dm})+1\Big]\Omega_{\rm dm}+\Big[2Q^2Y(k) +1\Big]\Omega_{\rm b}b}  \,,\label{EP final}
\end{align}
where we recall that $f_1=e^{-2Q\phi}$. This is the main result of this section.
Assuming baryons to be sub-dominant, $E_P$ reduces to
\begin{align} \label{eq:EP-CDE} 
    E_{P}&=1+Q_{\rm dm}\phi'-3Q_{\rm dm}Y(k)\frac{\Omega_{\rm dm}}{f_1}\frac{ Q+Q_{\rm dm}}{f_g}   \nonumber \\
    & =1+Q_{\rm dm}\phi'-3Q_{\rm dm}\tilde{\Omega}_{\rm dm}Y(k)\frac{ Q+Q_{\rm dm}}{f_g} \,.
\end{align}
Recall that $\tilde{\Omega}_{\rm dm} = \Omega_{\rm dm}/f_1$. As expected, $E_P\to 1$ when $Q_{\rm dm}\to 0$. We introduce the dimensionless variables 
\begin{equation}
x(N) =\frac{\phi'}{\sqrt{6}}; \quad y(N) = \frac{1}{H}\sqrt{\frac{V}{3f_1}}.
\end{equation}
Neglecting baryons, the effective dark energy and dark matter density parameters can then be expressed as
\begin{align}
    \tilde{\Omega}_{\rm DE, eff} & =\frac{\rho_{\rm DE, eff}}{3H^2f_1} = (1-6Q^2)x^2+y^2+2\sqrt{6}Qx     \,,  \\
    \tilde{\Omega}_{\rm dm}  &= 1- \tilde{\Omega}_{\rm DE, eff} = 1-(1-6Q^2)x^2-y^2 - 2\sqrt{6}Qx \label{eq: xy_tilde_dm} \,,
\end{align} 
and $E_P$ is given by
\begin{equation}
    E_{P} =1+ \sqrt{6}Q_{\rm dm}x-3Q_{\rm dm}\left[1-(1-6Q^2)x^2-y^2 - 2\sqrt{6}Qx\right]Y(k)\frac{ Q+Q_{\rm dm}}{f_g} \label{eq: E_P} \,.
\end{equation}
We now want to investigate how strong the couplings $Q_{\rm \rm dm}$ and $Q$ can be constrained by present and future surveys. To this end, two scenarios are considered: the CDE case in which only $Q_{\rm dm}$ is nonzero, and the DCDE case in which both $Q$ and $Q_{\rm dm}$ are nonzero.

The first scenario is interesting since it corresponds to the simplest non-trivial coupled model, in which $f_1=1$, consequently $Q=0$ and $\Omega_{\rm dm}=\tilde{\Omega}_{\rm dm}$. In addition, we further assume $a^2 m_{\phi}^2\ll k^2$, so that $Y\to 1$. This is expected, as the mass of the dark energy scalar field needs to be extremely small in order to drive the late-time accelerated expansion of the Universe.
 The main motivation to examine closely this particular case, besides its simplicity, is that it has been shown to provide a good fit to the recent DESI data (cf. Refs \cite{Chakraborty:2024xas, Gomez-Valent:2026ept, Babu:2026tfr, Ganguly:2026mgu, Li:2026xaz, Wang:2025znm}).
Under these assumptions, $E_P$ can be further simplified and becomes $k$-independent: 
\begin{align}
E_{P}&=1+Q_{\rm dm}\phi'-3\frac{\Omega_{\rm dm}Q_{\rm dm}^2}{f_g} \nonumber \\
   &= 1+\sqrt{6}Q_{\rm dm}x- 3\frac{(1-x^2-y^2)Q_{\rm dm}^2}{f_g}             \,.   \label{EP Quintessence} 
\end{align}
Since we expect the behavior of the model to be close to that of $\Lambda$CDM to match observations, we can adopt the $\Lambda$CDM-like approximation $f_g \approx \Omega_{\rm dm}^{\gamma}$ for the growth rate, with $\Omega_{\rm dm}$ evaluated using the background evolution of the CDE model and $\gamma=0.545$.

By further introducing the dimensionless coupling\footnote{Our definition of the coupling $\beta_{\rm dm}$ differs slightly from the more common convention $\beta_{\rm dm} = -Q_{\rm dm}$ used e.g. in \cite{Gomez-Valent:2026ept, Tudes:2024jpg}.} $\beta_{\rm dm} = -\frac{\sqrt{6}}{2}Q_{\rm dm}$, we finally obtain $E_P$ for the coupled dark energy scenario:
\begin{align}
    E_{P}(N) = 1 - 2 \beta_{\rm dm} x - 2 \beta_{\rm dm}^2 (1 - x^2 - y^2)^{1-\gamma}  
    \label{eq:epfin}   \,.
\end{align}
The evolution of $x$ and $y$ depends on the shape of the potential. For generality, we employ the analytical approximate solutions for $x$ and $y$ valid around the present time derived in Ref.~\cite{Tudes:2024jpg}  for a linear potential $V\approx V_0-V_0\mu(\phi-\phi_0)$ (where $\mu$ is constant, and we set $\phi_0=0$ as before), which corresponds to the first-order Taylor expansion of a generic potential and satisfies $V'(N) = -V_0\mu\phi' = -V_0\mu\sqrt{6}x$. This yields 
\begin{align}
    x(N) &= \frac{1}{\sqrt{6}\,e^{3N}\tilde{E}(N)}
    \Bigg\{\frac{4}{\sqrt{6}}\beta_{\rm dm}\left[e^{3N_i}\tilde{E}(N_i)+\frac{\Omega_{\rm m0}}{\sqrt{\Omega_{\Lambda 0}}}\ln\left(\frac{e^{3N_i/2}(\tilde{E}(N_i)-\sqrt{\Omega_{\Lambda 0}})}{e^{3N/2}(\tilde{E}(N)-\sqrt{\Omega_{\Lambda 0}})}\right)\right]  \nonumber\\ 
    &+\mu\left[\frac{\Omega_{\rm m0}}{\sqrt{\Omega_{\Lambda 0}}}\ln\left(\frac{e^{3N/2}(\tilde{E}(N)-\sqrt{\Omega_{\Lambda 0}})}{e^{3N_i/2}(\tilde{E}(N_i)-\sqrt{\Omega_{\Lambda 0}})}\right)+e^{3N}\tilde{E}(N)-e^{3N_i}\tilde{E}(N_i)\right]\Bigg\} \label{eq:x} \,,
\end{align}
\begin{align}
    V (N) &=V_0\Bigg\{1-\frac{4}{\sqrt{6}}\beta_{\rm dm} N\mu+\mu\left(\frac{\frac{8}{\sqrt{6}}\beta_{\rm dm}-2\mu}{3\sqrt{\Omega_{\Lambda0}}}\right)\Bigg[\frac{\sqrt{\Omega_{\Lambda0}}}{\Omega_{\rm m0}}e^{3N_i}\tilde{E}(N_i)\left(\tilde{E}(N)-1\right)\nonumber \\
    &-\ln\Big(\frac{e^{3N_i/2}(\tilde{E}(N_i)-\sqrt{\Omega_{\Lambda0}})}{1-\sqrt{\Omega_{\Lambda0}}}\Big)
    +\tilde{E}(N)\ln\Big(\frac{e^{3N_i/2}(\tilde{E}(N_i)-\sqrt{\Omega_{\Lambda0}})}{e^{3N/2}(\tilde{E}(N)-\sqrt{\Omega_{\Lambda0}})}\Big)\Bigg]\Bigg\} \label{eq:y}  \,,
\end{align}
where $N_i$ is the initial time chosen at $N_i =-3$, $\Omega_{\Lambda0}=1-\Omega_{{\rm m}0}$, $V_0=3H_0^2\Omega_{\Lambda0}$, and $\tilde{E}(N) = \sqrt{\Omega_{\rm m0}e^{-3N} + 1-\Omega_{\rm m0}}$. The initial condition of $x$ is solved from the scaling solution during the $\phi$MDE phase, with $x_i(N=-3) = \frac{2}{3}\beta_{\rm dm}$.

Next, for the doubly coupled dark energy, where both $Q$ and $Q_{\rm dm}$ are active, we further define the dimensionless coupling $\beta = -\frac{\sqrt{6}}{2}Q$ associated with the universal coupling. Using the same approximation of $Y\to 1$ and $f_g \approx \tilde{\Omega}_{\rm dm}^{\gamma}$,  $E_P$ is given by
\begin{equation} \label{eq: Ep_xy_DCDE}
    E_{P} = 1-2\beta_{\rm dm}x-2\beta_{\rm dm}\left(\beta+\beta_{\rm dm}\right)\left[1-(1-4\beta^2)x^2-y^2 + 4\beta x\right]^{1-\gamma} \,.
\end{equation}
The quantities $x$ and $V$ in the doubly coupled model can be analytically approximated by:
\begin{align} \label{eq: x_QQdm}
x(N)&= \frac{e^{3N_i}\tilde{E}(N_i)}{e^{3N}\tilde{E}(N)}\,\frac{2(\beta+\beta_{\rm dm})}{\big[3-4\beta(\beta+\beta_{\rm dm})\big]}+\frac{1}{3(1-4\beta^2)}
\left[4\beta+\sqrt{\frac{3}{2}}\mu\right]
\left[1-\frac{e^{3N_i}\tilde{E}(N_i)}{e^{3N}\tilde{E}(N)}\right]  \nonumber \\
&+\frac{\Omega_{\rm m0}}{3(1-4\beta^2)\sqrt{1-\Omega_{\rm m0}}e^{3N}\tilde{E}(N)}
\left[2\beta_{\rm dm}-2\beta-\sqrt{\frac{3}{2}}\mu\right] \nonumber \\
&\qquad \qquad \qquad \times\ln\left[\frac{e^{3N_i/2}\left(\tilde{E}(N_i)-\sqrt{1-\Omega_{\rm m0}}\right)}
{e^{3N/2}\left(\tilde{E}(N)-\sqrt{1-\Omega_{\rm m0}}\right)}\right].
\end{align}
\begin{align} \label{eq: V_QQdm}
V(N) &= V_0\Bigg\{1-\frac{2\sqrt{6}\,\mu(\beta+\beta_{\rm dm})}{3(1-4\beta^2)}N+\frac{2\sqrt{6}\,\mu\,e^{3N_i}\tilde{E}(N_i)}{3\Omega_{\rm m0}}
\left[x(N_i)-\frac{4\beta+\sqrt{\frac{3}{2}}\mu}{3(1-4\beta^2)}\right]
\left[\tilde{E}(N)-1\right]  \nonumber\\
&+\frac{2\sqrt{6}\,\mu\left[2\beta_{\rm dm}-2\beta-\sqrt{\frac{3}{2}}\mu\right]}
{9(1-4\beta^2)\sqrt{\Omega_{\Lambda0}}}\times\Bigg[
\tilde{E}(N)\ln\Big(\frac{e^{3N_i/2}
\left[\tilde{E}(N_i)-\sqrt{\Omega_{\Lambda0}}\right]}{e^{3N/2}
\left[\tilde{E}(N)-\sqrt{\Omega_{\Lambda0}}\right]}\Big)
-\ln\Big(\frac{e^{3N_i/2}\left[\tilde{E}(N_i)-\sqrt{\Omega_{\Lambda0}}\right]}
{1-\sqrt{\Omega_{\Lambda0}}}\Big)\Bigg]\Bigg\}.
\end{align}
The derivation of Eqs. \eqref{eq: x_QQdm} and \eqref{eq: V_QQdm} is provided in App. \ref{app: xy}. Moreover, the phase-space system for the doubly coupled model including baryons is summarized in App. \ref{app: dynimicalapproach}. The expressions Eqs. \eqref{eq:epfin}-\eqref{eq: V_QQdm} will be  implemented in our forecasts. Next, we describe our forecasting methodology in Sec. \ref{sec: forecast} and provide the forecast results for the constant couplings $\beta$ and $\beta_{\rm dm}$ in Sec. \ref{sec: results}.

\section{Fisher forecast for the couplings} \label{sec: forecast}

In Ref.~\cite{EP}, it has been shown that the auto- and cross-matter linear power spectra for two galaxy populations (with the subscripts \(\text{B}, \text{F}\), for "bright" and "faint", respectively) up to order \(\lambda = \mathcal{H}/k\) can be written as
\begin{align}
P_{\F\F} & =\left(1+\beta_{\F}\mu^{2}\right)^{2}\frac{\beta_{\B}}{\beta_{\F}}S_{g}^{2}BP\,,\label{eq:spectra1}\\
P_{\B\B} & =\left(1+\beta_{\B}\mu^{2}\right)^{2}\frac{\beta_{\F}}{\beta_{\B}}S_{g}^{2}BP\,,\label{eq:spectra2}\\
P_{\F\B} & =\big[(1+\beta_{\F}\mu^{2})(1+\beta_{\B}\mu^{2})\nonumber \\
 & +i\lambda\mu\left(\tau_{1}+\mu^{2}\tau_{2}\right)]S_{g}^{2}BP\,,\label{eq:spectra3}
\end{align}
in which $\mu$ is the angle between the galaxy pair and the line of sight to the observer,  $\beta_{B,F}=f_g/b_{B,F}$ ($b_{B,F}$ being the time-dependent linear galaxy bias for each population), $P$ is
the power spectrum at $z=0$, and $B\equiv b_{\B}(z)b_{\F}(z)G^2(z)$,
with $G\equiv\delta_{\rm m}(z)/\delta_{\rm m0}$ denoting the linear growth factor.
The spectra are further multiplied by factors accounting for the Finger-of-God effect 
\begin{equation}
S_{g}^2(k,\mu,z)=e^{-\frac{1}{2}k^2\mu^2\sigma_{g}^{2}}\,,\label{eq:FoG}
\end{equation}
in which the damping strengths $\sigma_{g}$ are left
as free parameters in each redshift bin. Moreover, the parameters $\tau_{1}$
and $\tau_{2}$ appearing in the cross-spectrum are given by 
\begin{align}
\tau_{1} & =\beta_{\B}\alpha_{\B}-\beta_{\F}\alpha_{\F}-E_{P}(\beta_{\B}-\beta_{\F})\,,\label{eq:tau1}\\
\tau_{2} & =\beta_{\B}\beta_{F}(\alpha_{\B}-\alpha_{\F})\, ,\label{eq:tau2}
\end{align}
where we have defined for each galaxy population
\begin{equation}
    \alpha\equiv1-5s+\frac{5s-2}{\mathcal{H}r}-\frac{{{\mathcal H}'}}{\mathcal{H}}+f^{{\rm evol}}\,.
    \end{equation}
Here, $s$ is the magnification bias, accounting for the fact that surveys are flux limited, and $f^{{\rm evol}}$ is the evolution bias, encoding the evolution of galaxies. Here we simply set \(f^{\text{evol}} = 0\), as in our previous analysis~\cite{EP, Zheng:2025oiq}. Moreover, we assume in the following that $\alpha_{B,F}$ can be accurately constrained, as both the magnification and the evolution bias can be measured from the average number of galaxies in a given sample \cite{Maartens:2021dqy,Bonvin:2023jjq}. By grouping the power spectra Eqs. \eqref{eq:spectra1}-\eqref{eq:spectra3} into polynomials of $\mu^a\lambda^b$, one can identify measurable model-independent quantities.
From $\tau_1$ in Eq.~\eqref{eq:tau1}, we see then explicitly that $E_{P}$ can only be measured if the two galaxy populations have
different biases. This is the fundamental reason why we need two distinct source populations.

The spectra in Eqs, \eqref{eq:spectra1}-\eqref{eq:spectra3} encode also the Alcock-Paczynski (AP) effect through $k$ and $\mu$, relating to the reference cosmology through
\begin{equation}
    k = \alpha_{\rm AP}k_r; \quad \mu = \frac{\mu_rh}{\alpha_{\rm AP}}; \quad \alpha_{\rm AP}=\frac{1}{d}\sqrt{\mu_r^2(h^2d^2-1)+1} ,
\end{equation}
with $k_r$ and $\mu_r$ evaluated in the reference cosmology. These expressions depend
on $h\equiv E/E_{r}$ and on $d\equiv L_{A}/L_{Ar}$, in which $E(z)\equiv H(z)/H_{0}$
is the dimensionless Hubble function and $L_{A}(z)\equiv H_{0}D_{A}(z)$
is the dimensionless comoving angular diameter distance. Since $d$
is degenerate with the power spectrum at the linear level, we will
consider the combination $h_{d}\equiv hd$ as a parameter in our analysis.
In Ref.~\cite{EP}, a forecast analysis has been performed for a set of parameters
$\{h_{d},d,B,\sigma_{g},\beta_{\B},\beta_{\F},E_{P}\}$ at each redshift.
In addition, the power spectrum shape was parametrized in several wavebands
in the first redshift bin $z_{1}$ and evolved  with the free function
$B(z)$. The $k$-range was chosen to be $k\in(0.01-0.12)\,h/$Mpc with intervals
$\Delta k=0.01\,h/$Mpc, for a total of twelve $P(k)$ waveband parameters. Following Ref. \cite{EP}, we assume uninformative flat priors for all parameters except $d$, with a $1\%$ Gaussian prior on the latter, based on current constraints (see, e.g., the Pantheon+ supernova catalog \cite{Brout:2022vxf} and App. \ref{app:sn}), while fixing $B(z_1) =1$ to break the degeneracy with $P(k, z_1)$ (in other words, we only measure ratios of $B$ at different redshifts). 

Here, we follow the methodology of Refs. \cite{EP, Zheng:2025oiq}, with only slightly different survey specifications (see Sec. \ref{sec: specifications_fiducials}), to obtain constraints on $\{h_{d},d,B,\sigma_{g},\beta_{\B},\beta_{\F},E_{P}\}$ within each redshift bin. We then project these redshift-dependent constraints onto our constant model parameters. For the CDE scenario, we project onto $\{\beta_{\rm dm}, \Omega_{\rm m0}\}$, while for the DCDE scenario, the parameters are instead $\{\beta, \beta_{\rm dm}, \Omega_{\rm m0}\}$. The couplings $\beta$ and $\beta_{\rm dm}$ enter both $E_P(z)$ (see Eqs. \eqref{eq:epfin} and \eqref{eq: Ep_xy_DCDE}) and the expansion rate $E(z)$ (see Eqs. \eqref{eq: APPC_E_CDE} and \eqref{eq: E_exact}), with the latter modifying $h_d$ and $d$ through the AP effect. We therefore project the  marginalised constraints on $\{E_P, h_d, d\}$ in each redshift bin onto the CDE and DCDE model parameters, respectively\footnote{In principle, $\beta$ and $\beta_{\rm dm}$ also modify the clustering of matter by altering $f_g(z)$ and $G(z)$, and therefore enter $\beta_B, \beta_F$ and $B$. Indeed, one can approximate $f_g\approx\tilde{\Omega}_{\rm dm}^\gamma$ using the $\tilde{\Omega}_{\rm dm}$ given by Eq. \eqref{eq: xy_tilde_dm}, or employ a numerical fit for $f_g(\beta, \beta_{\rm dm})$ (see, e.g., the growth rate fitting function for the CDE scenario with a linear or exponential potential in Ref. \cite{Tudes:2024jpg}). However, the dependence of the galaxy biases on $\beta, \beta_{\rm dm}$ remains unknown, resulting in the dependence of $B$ and $\beta_{B,F}$ on the couplings being vague. For this reason, we perform the Fisher projection using only the marginalised constraints on $\{E_P, h_d, d\}$ as inputs.}. We summarize the details of the Fisher projection in App. \ref{app: Projection}.

\section{Survey specifications and fiducials} \label{sec: specifications_fiducials}
Here we consider three upcoming galaxy surveys and forecast how well they can constrain the coupling strengths: the DESI Bright Galaxy Sample (BGS) \cite{DESI:2016fyo}, a Euclid-like stage-IV H$\alpha$ spectroscopic survey \cite{Euclid:2024yrr}, and the SKA Phase 2 HI spectroscopic galaxy survey \cite{Bull:2015lja}. In addition, we compute the fiducial matter power spectrum at the lowest redshift bin, background quantities, and linear growth rate assuming a standard flat \(\Lambda\text{CDM}\) cosmology based on the \textit{Planck} 2018 results \cite{Planck:2018vyg}. To forecast the couplings, we also need to specify the initial time $N_i$ and the fiducial values of $\beta$, $\beta_{\rm dm}$, and $\Omega_{\rm m0}$ in the expressions of $x$ and $y$. Following Ref.~\cite{Tudes:2024jpg}, we adopt $N_i=-3$, deep in the matter-dominated era, where $x_i$ can be determined from the $\phi$MDE scaling solution (see App. \ref{app: xy}). Finally, we take the fiducial values
$\Omega_{\rm m0}=0.315$, $\beta = 0.1$, $\beta_{\rm dm}=0.2$, $V_0=3(1-\Omega_{\rm m0})\approx 2.055$ and $\mu=0.145$. 
The fiducial values for the couplings are larger than  current constrains obtained  by comparison with CMB and other cosmological data (see e.g. \cite{Gomez-Valent:2026ept, Chakraborty:2025syu, Gomez-Valent:2020mqn}). These analyses, however, assume a constant coupling throughout the entire cosmic evolution, while here we only consider late-time data.

For each survey, we split the full galaxy sample into two populations (bright and faint) within each redshift bin. This is done by choosing a flux threshold that divides the sample equally between the two populations. As in \cite{EP}, we choose the fiducial biases to be $b_{g, B}=b_g+0.5$ and $b_{g, F}=b_g-0.5$, where $b_g$ is the mean galaxy bias of the full sample. This choice is consistent with the measurements reported in \cite{Bonvin:2023jjq}.

The remaining survey specifications for DESI BGS, Euclid, and SKA2 are identical to those in Ref. \cite{Zheng:2025oiq}.  We list them explicitly in Tables~\ref{tab:DESI_specs}--\ref{tab:SKA2_specs}. These include the redshift bins, survey volumes, galaxy number densities, and average galaxy biases, as well as the treatment of galaxy redshift uncertainties (associated with $\sigma_g$) and the prescriptions for magnification and evolution biases.

The forecasting results presented in the next section are based on the aforementioned survey specifications.
\begin{table*}
\caption{Survey specifications for DESI BGS covering the redshift range $0 \leq z \leq 0.5$ in bins of width $\Delta z = 0.1$. Here we list the central redshift of each bin. In this table and hereafter, we express the galaxy number density $n_g$ in units of $10^{-3}\,(h/\mathrm{Mpc})^3$, assuming $h = 0.67$, and the survey volumes $V$ in $(\mathrm{Gpc}/h)^3$. The fiducial bias are assumed as $b_{g,B}, b_{g,F} = b_g\pm 0.5$. The galaxy number density $n_g$ and the magnification bias $s_B, s_F$, are obtained from \cite{Maartens:2021dqy}.  The observed absolute magnitude limit is chosen to be \( m_c = 19.5 \). The evolution bias, $f_{\rm evol}$, is set to zero.} \label{tab:DESI_specs}. 
\centering
\renewcommand{\arraystretch}{1.5} 
\begin{tabular}{ccccccccccc}
$z$ & $V$ & $n_g$ & $b_g$ & $s_B$ & $s_F$ & $\beta_B$ & $\beta_F$ & $\alpha_B$ & $\alpha_F$  & $B$ \\
\hline\hline
0.05 & 0.04 & 56.7 & 1.38 & 0.243 & 0.11 & 0.297 & 0.634 & -17 & -30.1 & 1.00 \\
0.15 & 0.23 & 16.9 & 1.45 & 0.473 & 0.219 & 0.314 & 0.645 & 0.938 & -7.16 & 1.01 \\
0.25 & 0.58 & 5.54& 1.53 & 0.779 & 0.441 & 0.326 & 0.642 & 5.66 & -0.544 & 1.02 \\
0.35 & 1.04 & 1.42 & 1.61 & 1.21 & 0.812 & 0.334 & 0.635 & 8.9 & 3.93 & 1.03 \\
0.45 & 1.55 & 0.25& 1.7 & 1.81 & 1.37 & 0.338 & 0.619 & 11.8 & 7.73 & 1.05 \\
\hline\hline
\end{tabular}
\end{table*}

\begin{table*}
\caption{\label{tab:Euclid_specs} Survey specifications for a Euclid-like spectroscopic survey, covering the redshift range $0.9 \leq z \leq 1.8$ in four redshift bins with boundaries at $z = 0.9, 1.1, 1.3, 1.5, 1.8$. The quantities below are derived from Refs. \cite{Euclid:2019clj, Maartens:2021dqy}.}
\centering
\renewcommand{\arraystretch}{1.5} 
\begin{tabular}{cccccccccccc}
$z$ & $V$ & $n_g$ & $b_g$ & $s_B$ & $s_F$ & $\beta_B$ & $\beta_F$ & $\alpha_B$ & $\alpha_F$  & $B$ \\
\hline\hline
1.0 & 7.94 & 0.686 & 1.46 & 1.04 & 0.904 & 0.447 & 0.912 & 0.664 & 0.344   & 1 \\
1.2& 9.15 & 0.558 & 1.61 & 1.08 & 0.974 & 0.428 & 0.813 & 0.122 & -0.01  & 1.05 \\
1.4 & 10.1 & 0.421 & 1.75  & 1.1 & 1.03 & 0.41 & 0.738 & -0.32 & -0.36   & 1.08 \\
1.65 & 16.2 & 0.261 & 1.9 & 1.12 & 1.07 & 0.392 & 0.672 &  -0.762 & -0.755 & 1.07 \\
\hline\hline
\end{tabular}
\end{table*}

\begin{table*}
\caption{SKA2 survey specifications covering the redshift range $0.2 \leq z \leq 0.8$ in bins of width $\Delta z = 0.1$. The quantities below are derived from Refs.~\cite{Bull:2015lja, Maartens:2021dqy}.}
\centering
\renewcommand{\arraystretch}{1.5} 
\begin{tabular}{cccccccccccc} \label{tab:SKA2_specs}
$z$ & $V$ & $n_g$ & $b_g$ & $s_B$ & $s_F$ & $\beta_B$ & $\beta_F$ & $\alpha_B$ & $\alpha_F$  & $B$ \\
\hline\hline
0.25 & 1.2 & 121. & 0.674 & 0.364 & -0.026 & 0.563 & 3.8 & -1.96 & -9.12  & 1 \\
0.35 & 2.1 & 71.8 & 0.73 & 0.469 & 0.0235 & 0.573 & 3.06 & -0.35 & -5.91  & 1.25 \\
0.45 & 3.09 & 43.6 & 0.79 & 0.541 & 0.139 & 0.576 & 2.56 & 0.157 & -3.52  & 1.49 \\
0.55 & 4.11 & 26.8 & 0.854 & 0.608 & 0.276 & 0.573 & 2.19 & 0.389 & -1.92  & 1.72 \\
0.65 & 5.11 & 17. & 0.922 & 0.68 & 0.412 & 0.565 & 1.9 & 0.523 & -0.93  & 1.95 \\
0.75 & 6.06 & 10.9 & 0.996 & 0.755 & 0.539 & 0.554 & 1.67 & 0.576 & -0.346  & 2.19 \\
\hline\hline
\end{tabular}
\end{table*}

\section{Results and discussion} \label{sec: results}

\begin{table*} 
\caption{Absolute $1\sigma$  marginalized errors for DESI BGS, assuming a fiducial galaxy bias $b_{g,B},b_{g,F}=b_g\pm0.5$. We present the model-independent constraints on ${h_d,d,E_P}$ in each redshift bin, together with their projections onto the CDE and DCDE models. The model parameters ${\beta,\beta_{\rm dm},\Omega_{m0}}$ are assumed to be constant along all the redshift bins. Fiducial values are chosen to be $\beta=0.1, \beta_{\rm dm}=0.2, \Omega_{\rm m0}=0.315$.}
\begin{ruledtabular}   \label{tab: result_DESI}
\renewcommand{\arraystretch}{1.15}
\setlength{\tabcolsep}{6pt}
\begin{tabular}{
    c
    D{.}{.}{1.3}
    D{.}{.}{1.3}
    D{.}{.}{2.1}
    !{\hspace{10pt}}
    cc
    !{\hspace{10pt}}
    ccc
}
 & \multicolumn{3}{c}{Model-independent constraints}
 & \multicolumn{2}{c}{CDE projection}
 & \multicolumn{3}{c}{DCDE projection} \\ [4pt]
  $z$
 & \multicolumn{1}{c}{$h_d$}
 & \multicolumn{1}{c}{$d$}
 & \multicolumn{1}{c}{$E_P$}
 & $\beta_{\rm dm}$
 & $\Omega_{\rm m0}$
 & $\beta$
 & $\beta_{\rm dm}$
 & $\Omega_{\rm m0}$ \\
\colrule
0.05 & 0.068 & 0.011 & 1.7
 & \multirow{5}{*}{0.96}
 & \multirow{5}{*}{0.29}
 & \multirow{5}{*}{3.1}
 & \multirow{5}{*}{2.7}
 & \multirow{5}{*}{0.51} \\
0.15 & 0.049 & 0.0099 & 3.3 & & & & & \\
0.25 & 0.042 & 0.0098 & 4.2 & & & & & \\
0.35 & 0.04 & 0.0097 & 6.7 & & & & & \\
0.45 & 0.046  & 0.0098 & 1\rlap{$5$} & & & & & \\
\end{tabular}
\end{ruledtabular}
\end{table*}

\begin{table*} 
\caption{Same as Table. \ref{tab: result_DESI}, but for the Euclid spectroscopic survey. }
\begin{ruledtabular}  \label{tab: result_Euclid}
\renewcommand{\arraystretch}{1.15}
\setlength{\tabcolsep}{6pt}
\begin{tabular}{
    c
    D{.}{.}{1.3}
    D{.}{.}{1.3}
    D{.}{.}{2.1}
    !{\hspace{10pt}}
    cc
    !{\hspace{10pt}}
    ccc
}
 & \multicolumn{3}{c}{Model-independent constraints}
 & \multicolumn{2}{c}{CDE projection}
 & \multicolumn{3}{c}{DCDE projection} \\ [4pt]
  $z$
 & \multicolumn{1}{c}{$h_d$}
 & \multicolumn{1}{c}{$d$}
 & \multicolumn{1}{c}{$E_P$}
 & $\beta_{\rm dm}$
 & $\Omega_{\rm m0}$
 & $\beta$
 & $\beta_{\rm dm}$
 & $\Omega_{\rm m0}$ \\
\colrule
1.0 & 0.014 & 0.0084 & 2.5
 & \multirow{5}{*}{0.69}
 & \multirow{5}{*}{0.15}
 & \multirow{5}{*}{2.2}
 & \multirow{5}{*}{2.0}
 & \multirow{5}{*}{0.24} \\
1.2 & 0.014 & 0.0084 & 2.9 & & & & & \\
1.4 & 0.015 & 0.0084 & 3.6 & & & & & \\
1.65 & 0.014  & 0.0082 & 4.1 & & & & & \\
\end{tabular}
\end{ruledtabular}
\end{table*}

\begin{table*} 
\caption{Same as Table. \ref{tab: result_DESI}, but for the SKA2.}
\begin{ruledtabular}  \label{tab: result_SKA2}
\renewcommand{\arraystretch}{1.15}
\setlength{\tabcolsep}{6pt}
\begin{tabular}{
    c
    D{.}{.}{1.3}
    D{.}{.}{1.3}
    D{.}{.}{2.1}
    !{\hspace{10pt}}
    cc
    !{\hspace{10pt}}
    ccc
}
 & \multicolumn{3}{c}{Model-independent constraints}
 & \multicolumn{2}{c}{CDE projection}
 & \multicolumn{3}{c}{DCDE projection} \\ [4pt]
  $z$
 & \multicolumn{1}{c}{$h_d$}
 & \multicolumn{1}{c}{$d$}
 & \multicolumn{1}{c}{$E_P$}
 & $\beta_{\rm dm}$
 & $\Omega_{\rm m0}$
 & $\beta$
 & $\beta_{\rm dm}$
 & $\Omega_{\rm m0}$ \\
\colrule
0.25 & 0.0019 & 0.012 & 0.089
 & \multirow{5}{*}{0.044}
 & \multirow{5}{*}{0.011}
 & \multirow{5}{*}{0.43}
 & \multirow{5}{*}{0.26}
 & \multirow{5}{*}{0.076} \\
0.35 & 0.002 & 0.01 & 0.082 & & & & & \\
0.45 & 0.0023 & 0.0091 & 0.079 & & & & & \\
0.55 & 0.0027  & 0.0085 & 0.15 & & & & & \\
0.65 & 0.0033 & 0.0084 & 0.2 & & & & & \\
0.75 & 0.004  & 0.0083 & 0.25 & & & & & \\
\end{tabular}
\end{ruledtabular}
\end{table*}

\begin{table*} 
\caption{Same as Table. \ref{tab: result_SKA2}, but with ten times as many galaxies.}
\begin{ruledtabular}  \label{tab: result_10SKA2}
\renewcommand{\arraystretch}{1.15}
\setlength{\tabcolsep}{6pt}
\begin{tabular}{
    c
    D{.}{.}{1.3}
    D{.}{.}{1.3}
    D{.}{.}{2.1}
    !{\hspace{10pt}}
    cc
    !{\hspace{10pt}}
    ccc
}
 & \multicolumn{3}{c}{Model-independent constraints}
 & \multicolumn{2}{c}{CDE projection}
 & \multicolumn{3}{c}{DCDE projection} \\ [4pt]
  $z$
 & \multicolumn{1}{c}{$h_d$}
 & \multicolumn{1}{c}{$d$}
 & \multicolumn{1}{c}{$E_P$}
 & $\beta_{\rm dm}$
 & $\Omega_{\rm m0}$
 & $\beta$
 & $\beta_{\rm dm}$
 & $\Omega_{\rm m0}$ \\
\colrule
0.25 & 0.001 & 0.014 & 0.074
 & \multirow{5}{*}{0.0081}
 & \multirow{5}{*}{0.003}
 & \multirow{5}{*}{0.21}
 & \multirow{5}{*}{0.12}
 & \multirow{5}{*}{0.038} \\
0.35 & 0.00073& 0.013 & 0.042 & & & & & \\
0.45 & 0.00066 & 0.01 & 0.03 & & & & & \\
0.55 & 0.00064  & 0.0088 & 0.0093 & & & & & \\
0.65 & 0.0011 & 0.0084 & 0.053 & & & & & \\
0.75 & 0.0013  & 0.0083 & 0.073 & & & & & \\
\end{tabular}
\end{ruledtabular}
\end{table*}

\begin{figure*} 
\centering
\includegraphics[width=1\textwidth]{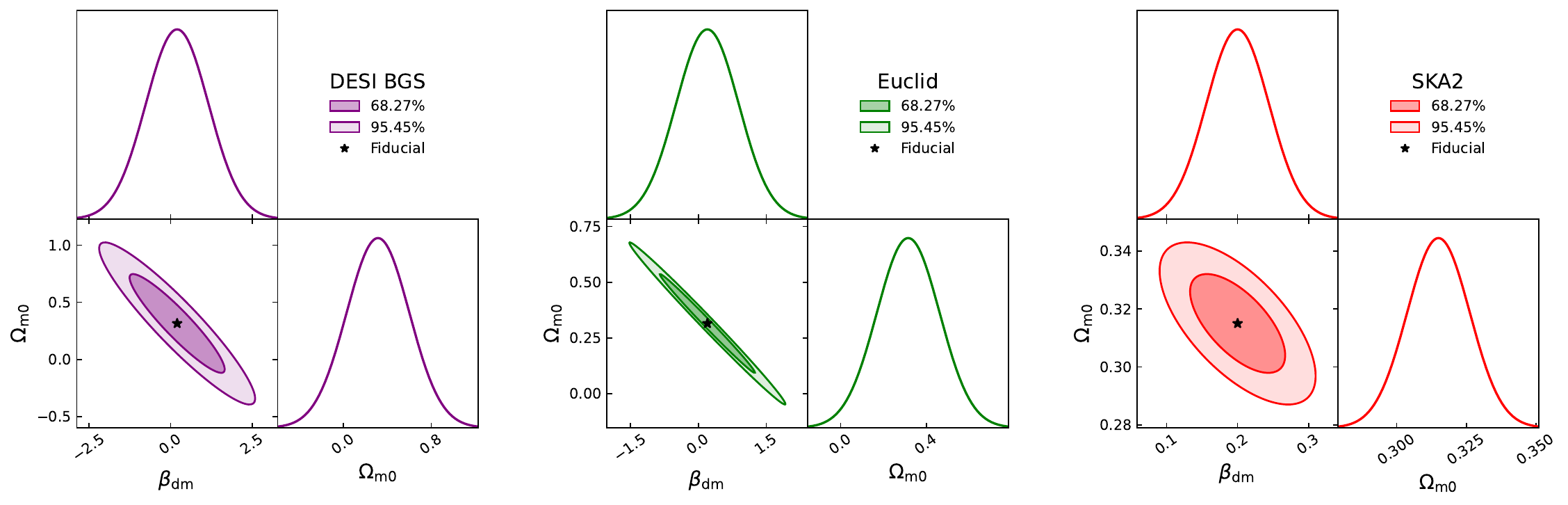}
\caption{Forecast constraints on $\beta_{\rm dm}$ and $\Omega_{\rm m0}$ for the CDE model. Here we show both one-dimensional posterior distributions of these two parameters, together with their 2D contour plots, within the 1$\sigma$ and 2$\sigma$ confidence region. The fiducial values are indicated as a star.}
    \label{CDE_triangle} 
\end{figure*}

We present our forecasting results in Tables \ref{tab: result_DESI}--\ref{tab: result_10SKA2} for DESI, Euclid, SKA2, and, just to illustrate the effect of the number density, a survey mimicking SKA2 but with ten times as many galaxies. All the results shown in these tables are absolute errors obtained after marginalising over all other parameters. No priors are taken into account except those on $d$, as illustrated in Sec. \ref{sec: forecast}. In each table, we present both model-independent constraints on $\{h_d, d, E_P\}$ within each redshift bin and model-independent constraints on the parameters of the CDE and DCDE models obtained by projection. We emphasize that here we only employ the forms of $E_P$ and the expansion rate $E$ (which enters $h_d$ and $d$) corresponding to the models onto which the constraints are projected, while the underlying analysis remains general as far as the cosmological model is concerned. By doing so, we expect our results to be more robust than those obtained from analyses carried out assuming specific models, at the price of possibly weaker constraints. Moreover, these constraints are fully obtained from the late-time Universe, in contrast to  tighter constraints obtained including CMB measurements and assuming constant couplings throughout the evolution. 

Among the three surveys, SKA2 could provide the tightest constraints on $h_d$, $d$ and $E_P$, consistent with the results reported in Refs. \cite{EP, Zheng:2025oiq}, thanks to its large number of detected galaxies. Projecting these constraints onto the CDE model, we obtain relative 1$\sigma$ errors of $22\%$ and $3.5\%$ on $\beta_{\rm dm}$ and $\Omega_{\rm m0}$, respectively. Let us first discuss the constraints on $\beta_{\rm dm}$, which directly reflects EP violation. From Eq.~\eqref{eq:epfin}, the uncertainty of $\beta_{\rm dm}$ inferred from $E_P$ at a single redshift bin, while keeping $\Omega_{\rm m0}$ fixed, can be approximated by
\begin{equation}
\sigma_{\beta_{\rm dm}}\approx \Big|\frac{1}
{\partial E_P/\partial\beta_{\rm dm}}\Big|_{\rm fid}\sigma_{E_P}  \,.
\end{equation}
For SKA2, $|\partial E_P/\partial\beta_{\rm dm}|_{\rm fid} \sim 0.9-1.12$ across the six redshift bins, evaluated at the fiducial. Assuming independent bins, the $1\sigma$ uncertainty of $\beta_{\rm dm}$ from $E_P$ alone is approximately $\sigma_{\beta_{\rm dm}}\approx0.044$, corresponding to a relative 1$\sigma$ uncertainty of $22.1\%$. This estimate is comparable to our forecast constraint of $\sigma_{\beta_{\rm dm}}=0.044$, as listed in Table~\ref{tab: result_SKA2}, despite neglecting the dependence of $h_d$ and $d$ on $\beta_{\rm dm}$ and correlation between bins. We also verified that, when including this, accounting for correlations between bins, and fixing $\Omega_{\rm m0}$, one can obtain $\sigma_{\beta_{\rm dm}}=0.034$. These results indicate that, for the upcoming Stage IV surveys considered here, $E_P$ measured from SKA2 can provide useful constraints on $\beta_{\rm dm}$ within the CDE model. The constraints on $\beta_{\rm dm}$ can be further improved at the cost of reducing model independence. For instance, we explored the case in which the power spectrum is fixed at fiducial, while leaving the remaining parameters free to vary. In this case, the relative $1\sigma$ error on $\beta_{\rm dm}$ for SKA2 is reduced to $19\%$.

On the other hand, $\beta_{\rm dm}$ not only affects perturbations by modifying the Euler equation and the growth rate, but also modifies the background expansion rate $E$. This introduces a degeneracy between the two parameters and increases the marginalised uncertainty on $\Omega_{\rm m0}$ when projecting the constraints on $h_d$ and $d$ onto the model parameters. Indeed, the correlation coefficients between $\beta_{\rm dm}$ and $\Omega_{\rm m0}$ are $-0.93$, $-0.99$, and $-0.65$ for DESI, Euclid, and SKA2, respectively. We also repeat the analysis but with a $\Lambda$CDM expansion, and the correlation coefficients become $-0.0094$, $0.0019$, $0.056$ for the three surveys. For SKA2, which is the most promising survey in this paper, the marginalised 1$\sigma$ error on $\beta_{\rm dm}$ remains approximately the same at $0.044$, but the absolute error on $\Omega_{\rm m0}$ decreases to $0.0032$. 

We also present in Fig.~\ref{CDE_triangle} the two-dimensional confidence contours between $\beta_{\rm dm}$ and $\Omega_{\rm m0}$ for all three surveys, together with their one-dimensional marginalised posterior distributions. In all three cases, the two parameters are negatively correlated. Moreover, the eigenvector $(c_1, c_2)$ associated with the smallest eigenvalue of the covariance matrix gives the linear combination $C = c_1\beta_{\rm dm}+c_2\Omega_{\rm m0}$ with the minimum variance. This eigenvector points along the short axis of the confidence ellipse. We find $(c_1, c_2)\simeq{(0.27, 0.96), (0.21, 0.98), (0.17, 0.99)}$ for DESI, Euclid, and SKA2, respectively. The corresponding relative $1\sigma$ uncertainties on $C$ are $29\%$, $4.6\%$, $2.5\%$. Thus, despite the large uncertainties on the individual parameters, this particular combination can be much better constrained, especially for Euclid, as can be seen directly from Fig.~\ref{CDE_triangle}. 

For the DCDE model, one can further constrain the universal coupling $\beta$, as it enters $E_P$ differently from $\beta_{\rm dm}$. The constraints are weaker, as expected, since we now have an additional parameter to vary. In the most tightly constrained case of SKA2, we obtain 1$\sigma$ relative errors of $430\%$, $130\%$, $24\%$ for $\beta$, $\beta_{\rm dm}$ and $\Omega_{\rm m0}$, respectively. Note that even though $\beta$ also alters the expansion and the growth rate, only $\beta_{\rm dm}$ is associated with the violation of the EP. To improve the constraints on $\beta$, one can add more cosmological probes, such as the CMB measurements from the early Universe.

We also explored the effect of increasing the galaxy number density. For instance, increasing the SKA2 galaxy number density by a factor of ten could substantially improve the projected constraints, as listed in Table \ref{tab: result_10SKA2}. For CDE, the 1$\sigma$ uncertainties decrease to $\sigma_{\beta_{\rm dm}}=0.0081$ and $\sigma_{\Omega_{\rm m0}}=0.003$, corresponding to relative errors of $4.1\%$ and  $0.95\%$, respectively. For DCDE, the 1$\sigma$ uncertainties decrease to $\sigma_\beta=0.21$, $\sigma_{\beta_{\rm dm}}=0.12$, and $\sigma_{\Omega_{\rm m0}}=0.038$, improving by roughly a factor of two. Moreover, constraints on $E_P$ can be significantly improved, reaching $3\%$ at $z=0.45$.

Lastly, this methodology can be extended to other EP-violating models or to alternative shapes of the scalar-field potential. Once the corresponding $E_P$ and expansion rate $E$ are identified, one can directly employ the framework presented here to investigate constraints on those model parameters.

\section{Conclusion}

In this paper, we studied model-independent cosmological tests of the weak equivalence principle in the context of a doubly coupled dark energy model, in which a scalar field couples simultaneously to gravity (associated with $\beta$) and dark matter (associated with $\beta_{\rm dm}$). When $\beta$ vanishes, this model reduces to coupled quintessence. We derived the background and linear perturbation equations in this DCDE scenario and showed that the effective potential is modified by a Yukawa correction. Additionally, both couplings modify the background expansion, the growth equation, the Euler equation, and therefore enter $E_P$. As expected, $E_P$ reduces to unity when $\beta_{\rm dm}$ is switched off, even with a non-vanishing universal coupling. Employing a dynamical system approach, we then obtained analytical expressions for $E_P$ in terms of the two couplings, assuming a linear potential. 

We then investigated whether these couplings, assumed here to be constant during the relevant late-time epoch, could be detected by next-generation cosmological surveys, namely DESI, Euclid and SKA2. To preserve robustness, we projected the model-independent constraints on $\{E_P, h_d, d\}$, obtained similarly to our previous analyses \cite{EP, Zheng:2025oiq} using two different galaxy populations, onto the CDE and DCDE model parameters. This could give us, for the first time, constraints on $\beta_{\rm dm}$ derived from direct cosmological tests of EP violations.

To summarise our results, our forecasts reveal the potential of $E_P$ measurements from SKA2 to constrain $\beta_{\rm dm}$ within the CDE model. However, obtaining tight constraints remains challenging for DESI and Euclid, as well as for the parameters of the DCDE model. This is due both to the limited sensitivity of $E_P$ to the couplings and to degeneracies between the couplings and other parameters. Increasing the galaxy number density could significantly help improve the constraints. 

Our results suggest several directions for further investigation. First, one could explore how the constraints depend on more complex shapes of the scalar-field potential. This requires to solve for $x$ and $y$ numerically. Second, constraints on $E_P$ could be improved by selecting galaxy populations with a larger bias difference, as shown in Ref.~\cite{EP}. One could also investigate whether extending the analysis to more than two galaxy populations provides further improvements. Finally, adding more cosmological probes could help reduce the degeneracy between $\beta$ and $\beta_{\rm dm}$ and improve constraints on the latter in the DCDE model, and therefore worth investigating this model through a full MCMC analysis. We leave these investigations for future work. 

\vspace{0.5cm}
\noindent {\bf Acknowledgments }
\newline
\newline
The authors thank C. Bonvin and S. Castello for useful comments. 
LA acknowledges support by DFG  under Germany's Excellence Strategy EXC 2181/1 - 390900948 (the Heidelberg STRUCTURES Excellence Cluster) and under Project  554679582 "GeoGrav: Cosmological Geometry
and Gravity with nonlinear physics" and  by Tamkeen under the NYUAD Research Institute grant CG014.

\appendix

\section{Yukawa corrections to the effective gravitational potentials}  \label{app:proof}

From Eqs. \eqref{Poisson phi}, \eqref{Poisson psi} and \eqref{Euler dm old}, the effective potential for dark matter particles $\hat\Psi_{\rm dm}\equiv \Psi+Q_{\rm dm}\varphi$
is given by 
\begin{align}
\hat{\Psi}_{\rm dm} & \equiv \Psi+Q_{\rm dm}\varphi      \nonumber \\
&= -\frac{3\lambda^2}{2}\Big[\big(2QY(k)( Q+Q_{\rm dm})+1\big)\tilde{\Omega}_{\rm dm}\delta_{\rm dm}+\big(2Q^2Y(k)+1\big)\tilde{\Omega}_{\rm b}\delta_{\rm b}\Big] \nonumber \\
& \quad -\frac{3\lambda^2}{2}\Big[2Q_{\rm dm}Y(k)(Q+Q_{\rm dm})\tilde{\Omega}_{\rm dm}\delta_{\rm dm}+ 2Y(k)QQ_{\rm dm}\tilde{\Omega}_{\rm b}\delta_{\rm b}\Big]   \nonumber \\
&= -\frac{3\lambda^2}{2}\left\{ \left[1+2Y(Q+Q_{\rm dm})^2\right] \tilde{\Omega}_{\rm dm}\delta_{\rm dm} +  \left[ 1+2YQ(Q+Q_{\rm dm})\right]\tilde{\Omega}_{\rm b}\delta_{\rm b}     \right\}   \label{eq: yukawapsi} \,.
\end{align}
Following Ref. \cite{Amendola:2003wa}, the density contrast for a dark matter particle in the DCDE model,  located in empty space at the origin, is given by
\begin{equation}
    \tilde{\Omega}_{\rm dm}\delta_{\rm dm} = \frac{\bar{\rho}_{\rm dm}}{\rho_{\rm crit}}\frac{\rho_{\rm dm}-\bar{\rho}_{\rm dm}}{\bar{\rho}_{\rm dm}} = \frac{\delta\rho_{\rm dm}}{\rho_{\rm crit}} = \frac{\delta\rho_{\rm dm}}{3f_1H^2} \label{eq: Odm_ddm} \,.
\end{equation}
On the other hand we know that
\begin{equation}
    m_{\rm dm}(a)=m_{\rm dm,0}e^{Q_{\rm dm}\left[\phi(a)-\phi_0\right]} = a^3\int d^3\textbf{x}\delta\rho_{\rm dm}(\textbf{x}) \,.
\end{equation}
Solving this gives, 
\begin{equation}
    \delta\rho_{\rm dm}(\textbf{x})=\frac{m_{\rm dm}}{a^3}\delta^{(3)}_D(\textbf{x})   \,,
\end{equation}
with $\delta^{(3)}_D(\textbf{x})$ the Dirac delta function satisfying 
$\int d^3\textbf{x}\,\delta^{(3)}_D(\textbf{x})=1$. Transforming to Fourier space we obtain
\begin{equation}
    \delta\rho_{\rm dm}(\textbf{k})=\frac{m_{\rm dm}}{a^3} \,.
\end{equation}
Plugging back into Eq. \eqref{eq: Odm_ddm} 
\begin{equation}
    \frac{3}{2}\lambda^2\tilde{\Omega}_{\rm dm}\delta_{\rm dm} = \frac{3}{2}\lambda^2\frac{\delta\rho_{\rm dm}}{3f_1H^2}=\frac{m_{\rm dm}}{2f_1k^2a} \,,
\end{equation}
and $\hat{\Psi}_{\rm dm \leftarrow dm}$ (i.e. the effective potential sourced from this single dark matter particle) becomes
\begin{equation}
    \hat{\Psi}_{\rm dm \leftarrow dm} (\textbf{k}) = -\frac{m_{\rm dm}}{2af_1}\left[\frac{1}{k^2}+\frac{2(Q+Q_{\rm dm})^2}{k^2+a^2m_{\phi}^2}\right]  \,,
\end{equation}
and transforming to real space this gives (with $\textbf{r}=a\textbf{x}$ denoting the physical distance)
\begin{equation}
    \hat{\Psi}_{\rm dm \leftarrow dm} (\textbf{r}) = - \frac{Gm_{\rm dm}}{f_1r}\left[1+2(Q+Q_{\rm dm})^2e^{-m_{\phi}r}\right]  \,.
\end{equation}
Likewise, for a baryon particle located at the origin, its effective potential acting on a dark matter test particle is given by
\begin{equation}
    \hat{\Psi}_{\rm dm \leftarrow b} (\textbf{r}) = - \frac{Gm_{\rm b}}{f_1r}\left[1+2Q(Q+Q_{\rm dm})e^{-m_{\phi}r}\right]  \,.
\end{equation}
On the other hand, for a baryon test particle the effective potential is simply $\hat{\Psi}_{\rm b} = \Psi$, one can find this from Eq. \eqref{Euler b}. Consider a single baryon particle as the source, then the effective potential is
\begin{equation}
    \hat{\Psi}_{\rm b\leftarrow b} (\textbf{k}) = -\frac{m_{\rm b}}{2af_1}\left[\frac{1}{k^2}+\frac{2Q^2}{k^2+a^2m_{\phi}^2}\right]  \,,
\end{equation}
and transforming to real space this gives
\begin{equation}
    \hat{\Psi}_{\rm b\leftarrow b} (\textbf{r}) = - \frac{Gm_{\rm b}}{f_1r}\left[1+2Q^2e^{-m_{\phi}r}\right]  \,.
\end{equation}

\section{Approximate analytical solutions for $x(N)=\frac{\phi'}{\sqrt{6}}$ and $y(N) = \frac{1}{H}\sqrt{\frac{V}{3f_1}}$  in the doubly coupled model} \ \label{app: xy}

It has been shown in \cite{Tudes:2024jpg} that one can find approximate analytical solutions for $x$ and $y$ with a linear potential and with only the coupling $\beta_{\rm dm}$ switched on. Moreover, the comparison between the analytical and numerical solutions in \cite{Tudes:2024jpg} (see their Fig. 3) indicates that assuming a background close to $\Lambda$CDM is a good approximation for small couplings. In this appendix, we further derive the analytical solutions for $x$ and $y$ in DCDE with the same linear potential, while depending on both $\beta$ and $\beta_{\rm dm}$. In this appendix we also neglect the baryon contribution. For a dynamical system approach including the baryons, see the next App. \ref{app: dynimicalapproach}.

We first derive the expansion rate $E$ in this model. Following a similar derivation as in \cite{Tudes:2024jpg}, taking the time derivative of Eq. \eqref{eq: Friedmann_effctive} and plug in Eqs. \eqref{eq: continuity_dark_matter} and \eqref{eq: continuity_DE_eff}, we have
\begin{equation} 
\frac{H'}{H}= -\frac{3}{2}\left(1+\frac{p_{\rm DE,eff}}{3f_1H^2}\right) \,, \label{eq: H'H}
\end{equation}
where $p_{\rm DE,eff}$ is effective pressure of the dark energy component. Since we set $\kappa=1$, the energy-momentum tensor of the effective dark energy reads
\begin{equation}
    T_{\mu\nu}^{\rm DE, eff} = T_{\mu\nu}^{\phi}+\nabla_{\mu}\nabla_{\nu}f_1-g_{\mu\nu}\Box f_1 \,,
\end{equation}
and therefore 
\begin{equation}
    p_{\rm DE,eff}=\frac{1}{2}(1-6Q^2)f_1H^2\phi'^2-V+H^2\left[f_1''+\left(2+\frac{H'}{H}\right)f_1'\right] \,. \label{eq: peff_xy}
\end{equation}
Expressing the pressure in terms of $x$ and $y$, and using the background equations to cancel $f_1''$ and $H'/H$, we obtain
\begin{equation}
\frac{p_{\rm DE,eff}}{3f_1H^2}=\frac{4}{3}\beta^2+(1-4\beta^2)\left[\left(1-\frac{4}{3}\beta^2\right)x^2
-y^2-\frac{4}{3}\beta x\right]+\frac{4}{3}\beta\beta_{\rm dm}\left[1-(1-4\beta^2)x^2-y^2+4\beta x\right]-\frac{2\sqrt{6}}{3}\beta\frac{V_{,\phi}}{V}\,y^2 \,. \label{eq: E_DCM}
\end{equation}
It is easy to find that $\frac{p_{\rm DE,eff}}{3f_1H^2}=x^2-y^2$ if $\beta=0$. The expansion rate $E$ can be solved from Eq. \eqref{eq: H'H}:
\begin{equation} 
    E(N) = {\rm exp}\left[-\frac{3}{2}\int_0^N \Big(1+\frac{p_{\rm DE,eff}(N')}{3f_1(N')H^2(N')}\Big)dN'\right]  \, ,\label{eq: E_exact}
\end{equation}
by plugging in Eq. \eqref{eq: E_DCM}. We emphasize that this integral is an exact solution of $E$. Next, we proceed to solve for the approximate solutions for $x(N)$ and $y(N)$. Substituting $\phi'=\sqrt{6}x$; -$\frac{2}{\sqrt{6}}\beta=Q=-\frac{f_{1,\phi}}{2f_1}$;$ -\frac{2}{\sqrt{6}}\beta_{\rm dm}=Q_{\rm dm}=\frac{f_{2,\phi}}{2f_2}$; $V=V_0-V_0\mu\phi$, the Klein-Gordon equation \eqref{eq:KG_BG} becomes
\begin{equation}
x'+\left(3+\frac{H'}{H}\right)x+2\beta x^2
= \frac{1}{1-4\beta^2}\left[\frac{V_0\mu}{\sqrt{6}f_1H^2}+2\beta\left(2+\frac{H'}{H}\right)+\beta_{\rm dm}\tilde{\Omega}_{\rm dm}\right]  \,,
\label{eq: app_KG}
\end{equation} 
which is still exact. Assuming the background is close to $\Lambda$CDM (which indicates small $\beta, \beta_{\rm dm}$) and $w_{\rm DE,eff}=p_{\rm DE,eff}/\rho_{\rm DE,eff}\approx-1$, we have $f_1, f_2\approx1$;$ V_0\approx 3H_0^2(1-\Omega_{\rm m0})$ and
\begin{equation}
    \tilde{\Omega}_{\rm dm}\approx \frac{\Omega_{\rm m0}}{\Omega_{\rm m0}+(1-\Omega_{\rm m0})e^{3N}}; \quad \frac{H'}{H} \approx -\frac{3}{2}\tilde{\Omega}_{\rm dm} \approx -\frac{3\Omega_{\rm m0}}{2\left[\Omega_{\rm m0}+(1-\Omega_{\rm m0})e^{3N}\right]}; \quad H^2 \approx H_0^2(\Omega_{\rm m0}e^{-3N}+1-\Omega_{\rm m0}). \label{eq:app_KG_apps}
\end{equation}
In addition, within the low-redshift range considered in this paper, we can assume that the kinetic energy of the scalar field is much smaller than its potential energy, allowing us to neglect the term quadratic in $x$ in Eq. \eqref{eq: app_KG}, which now can be reduced to
\begin{equation} \label{eq: KG_simplified}
x'+\left(3-\frac{3\Omega_{\rm m0}e^{-3N}}{2\tilde{E}^2}\right)x
= \frac{1}{1-4\beta^2}\left[\frac{V_0\mu}{\sqrt{6}H_0^2\tilde{E}^2}+4\beta+(\beta_{\rm dm}-3\beta)\frac{\Omega_{\rm m0}e^{-3N}}{\tilde{E}^2}\right]  \,,
\end{equation}
recall that $\tilde{E}$ is previously defined as $\tilde{E}(N) = \sqrt{\Omega_{\rm m0}e^{-3N} + 1-\Omega_{\rm m0}}$. Eq. \eqref{eq: KG_simplified} has analytical solutions 
\begin{align}
x(N)&= \frac{e^{3N_i}\tilde{E}(N_i)}{e^{3N}\tilde{E}(N)}\,x_i(N_i)+\frac{1}{3(1-4\beta^2)}
\left[4\beta+\sqrt{\frac{3}{2}}\mu\right]
\left[1-\frac{e^{3N_i}\tilde{E}(N_i)}{e^{3N}\tilde{E}(N)}\right]  \nonumber \\
&+\frac{\Omega_{\rm m0}}{3(1-4\beta^2)\sqrt{1-\Omega_{\rm m0}}e^{3N}\tilde{E}(N)}
\left[2\beta_{\rm dm}-2\beta-\sqrt{\frac{3}{2}}\mu\right] \nonumber \\
&\qquad \qquad \qquad \times\ln\left[\frac{e^{3N_i/2}\left(\tilde{E}(N_i)-\sqrt{1-\Omega_{\rm m0}}\right)}
{e^{3N/2}\left(\tilde{E}(N)-\sqrt{1-\Omega_{\rm m0}}\right)}\right]. \label{eq: x__QQdm}
\end{align}
Using $V'(N) = -V_0\mu\phi' = -V_0\mu\sqrt{6}x$, one can further obtain
\begin{align}
V(N) &= V_0\Bigg\{1-\frac{2\sqrt{6}\,\mu(\beta+\beta_{\rm dm})}{3(1-4\beta^2)}N+\frac{2\sqrt{6}\,\mu\,e^{3N_i}\tilde{E}(N_i)}{3\Omega_{\rm m0}}
\left[x(N_i)-\frac{4\beta+\sqrt{\frac{3}{2}}\mu}{3(1-4\beta^2)}\right]
\left[\tilde{E}(N)-1\right]  \nonumber\\
&+\frac{2\sqrt{6}\,\mu\left[2\beta_{\rm dm}-2\beta-\sqrt{\frac{3}{2}}\mu\right]}
{9(1-4\beta^2)\sqrt{\Omega_{\Lambda0}}}\times\Bigg[
\tilde{E}(N)\ln\Big(\frac{e^{3N_i/2}
\left[\tilde{E}(N_i)-\sqrt{\Omega_{\Lambda0}}\right]}{e^{3N/2}
\left[\tilde{E}(N)-\sqrt{\Omega_{\Lambda0}}\right]}\Big)
-\ln\Big(\frac{e^{3N_i/2}\left[\tilde{E}(N_i)-\sqrt{\Omega_{\Lambda0}}\right]}
{1-\sqrt{\Omega_{\Lambda0}}}\Big)\Bigg]\Bigg\}.
\label{eq:V__QQdm}
\end{align}
One can verify that, in the $\beta=0$ limit, Eqs.~\eqref{eq: x__QQdm} and ~\eqref{eq:V__QQdm} reduce to Eqs.~\eqref{eq:x} and ~\eqref{eq:y}, as expected.

The final step is to find the initial condition $x_i\equiv x(N_i)$ as required by Eq.~\eqref{eq: x__QQdm}.  We can obtain from the fixed point that satisfies the conditions $x'=0$ and $y=0$. 
This is similar to deriving the $\phi$MDE phase in the CDE scenario, and Eq. \eqref{eq: xy_tilde_dm} reads 
\begin{equation} 
 \tilde{\Omega}_{\rm dm}(N_i) \equiv \frac{\rho_{\rm dm}(N_i)}{3f_1(N_i)H^2(N_i)} = 1-(1-6Q^2)x_i^2 - 2\sqrt{6}Qx_i =  1-(1-4\beta^2)x_i^2 + 4\beta x_i  \label{eq: scaling_Om} \,, 
\end{equation}
Note that Eq.~\eqref{eq: scaling_Om} follows from the exact constraint in the DCDE case. We no longer assume a background close to $\Lambda$CDM here or in the following derivation. For a constant non-vanishing density fraction of dark matter
\begin{equation}
    \frac{\tilde{\Omega}_{\rm dm}'}{\tilde{\Omega}_{\rm dm}} = \frac{\rho_{\rm dm}'}{\rho_{\rm dm}} - \frac{f_1'}{f_1} - 2\frac{H'}{H} = 0 \,,
\end{equation}
it is easy to find that 
\begin{equation}
 \frac{\rho_{\rm dm}'}{\rho_{\rm dm}} = Q_{\rm dm}\phi' -3 = -2\beta_{\rm dm}x_i -3; \quad \frac{f_1'}{f_1} = \frac{f_{1,\phi}\phi'}{f_1} = -2Q\phi' = 4\beta x_i, \quad 
\end{equation}
and therefore
\begin{equation}
    \frac{H'}{H} = -\frac{3}{2}-(\beta_{\rm dm}+2\beta)x_i   \label{eq: scaling_H'H} \,.
\end{equation}
Substituting Eqs. \eqref{eq: scaling_Om} and \eqref{eq: scaling_H'H} into Eq. \eqref{eq: app_KG}, we find
\begin{equation}
    x_i= \frac{2(\beta+\beta_{\rm dm})}{3-4\beta(\beta+\beta_{\rm dm})}  \,. \label{eq:xi_phiMDE}
\end{equation}
Setting $\beta=0$ recovers the CDE solution $x_i=2\beta_{\rm dm}/3$, as expected, while for $\beta_{\rm dm}=0$ the result agrees with Eq. 32 in Ref. \cite{Tsujikawa_2008}. 
In addition, the constant effective dark energy fraction during the scaling phase of the DCDE model is given by
\begin{equation}
    \tilde{\Omega}_{\rm DE,eff} = (1-4\beta^2)x_i^2 - 4\beta x_i= (1-4\beta^2)\left[\frac{2(\beta+\beta_{\rm dm})}{3-4\beta(\beta+\beta_{\rm dm})}\right]^2 - \frac{8\beta(\beta+\beta_{\rm dm})}{3-4\beta(\beta+\beta_{\rm dm})}  \,.
\end{equation}
We also discuss here a particular case. From Eq. (\ref{eq: scaling_H'H}), we see that, in this regime, we can identify the effective EoS
\begin{equation}
w_{\rm eff}\equiv-1-\frac{2}{3}\frac{H'}{H}=\frac{2}{3}(\beta_{\rm dm}+2\beta)x_i \,.
\end{equation}  Interestingly, for $\beta_{\rm dm}=-2\beta$, one can have an expansion rate identical to a pure matter-dominated universe, i.e. $w_{\rm eff}=0$, even in presence of dark energy, i.e. $x_i\not= 0$.
From the equivalent form of Eq. \eqref{Euler dm old} (again neglecting baryons)
\begin{equation}
\delta_{\rm dm}''+\left(1+\frac{\mathcal{H}'}{\mathcal{H}}-2\beta_{\rm dm}x\right)\delta_{\rm dm}'-\frac{3}{2}\left[1+\frac{4}{3}Y(k)(\beta+\beta_{\rm dm})^2\right]\left[1-(1-4\beta^2)x^2-y^2+4\beta x\right]\delta_{\rm dm}=0 \,,
\end{equation}
and plugging in this particular case with $y=0, Y(k)=1, \beta_{\rm dm}=-2\beta$ and $x=x_i$ it becomes
\begin{equation}
\delta_{\rm dm}''+\left(\frac{1}{2}-\frac{8\beta^2}{3+4\beta^2}\right)\delta_{\rm dm}'-\frac{9-4\beta^2}{2(3+4\beta^2)}\delta_{\rm dm}=0 \,.
\end{equation}
For $\beta^2<9/4$, the growing mode is given by $\delta_{\rm dm} \propto a$ and thus $f_g=1$, again the same as during the matter-dominated era in a $\Lambda$CDM Universe. The $E_P$, however, is given by
\begin{equation}
    E_P = \frac{9(1-4\beta^2)}{(3+4\beta^2)^2} \,,
\end{equation}
which could still deviate from unity. For this particular case, then, while the background and the linearly perturbed behavior are indistinguishable from uncoupled dark energy, the deviation from the equivalence principle can still be detected.

\section{Phase-space system for the doubly coupled model including baryons}
\label{app: dynimicalapproach}

In this appendix we rewrite the background equations in terms of the variables $x,y$ introduced in the main text and then identify some of the critical points.
Starting with the equation for $\Tilde{\Omega}_{\rm dm}$ with $\kappa=1$,

\begin{align}
1 & =\frac{\rho_{{\rm dm}}}{3f_{1}H^{2}}+\frac{\rho_{{\rm b}}}{3f_{1}H^{2}}+\frac{1}{2}(1-6Q^{2})\frac{\phi'{}^{2}}{3}+\frac{V(\phi)}{3f_{1}H^{2}}-\frac{f_{1,\phi}\phi'\,}{f_{1}}  \nonumber \\
 & =\tilde{\Omega}_{\rm dm}+\tilde{\Omega}_{\rm b}+(1-6Q^{2})x^{2}+y^{2}-4\beta x  \,.
\end{align}
 Introducing a third variable $z=\frac{1}{H}\sqrt{\frac{\rho_{\rm b}}{3f_1}}$ associated with the baryon density fraction, we get 
\begin{equation}
\tilde{\Omega}_{\rm dm}=1-(1-4\beta^{2})x^{2}-y^{2}+4\beta x -z^2 \,.
\end{equation}
From
\begin{equation}
\frac{H'}{H}=-\frac{3}{2}\left(1+\frac{p_{{\rm DE,eff}}}{3f_{1}H^{2}}\right)\,,\label{eq: H'H-1}
\end{equation}
we include baryons and get
\begin{align}
-\frac{2}{3}\frac{H'}{H} & =1+\frac{4}{3}\beta^{2}+(1-4\beta^{2})\left[\left(1-\frac{4}{3}\beta^{2}\right)x^{2}-y^{2}-\frac{4}{3}\beta x\right]+\frac{4}{3}\beta\beta_{{\rm dm}}\left[1-(1-4\beta^{2})x^{2}-y^{2}+4\beta x - z^2 \right]-\frac{2\sqrt{6}}{3}\beta\frac{V_{,\phi}}{V}\,y^{2} \,. \label{eq: H'H-1-1}
\end{align}
Putting now $V=V_{0}e^{-\mu\phi}$, we obtain an equation containing only $x,y,z$:
\begin{align}
-\frac{2}{3}\frac{H'}{H} & =1+\frac{4}{3}\beta^{2}+(1-4\beta^{2})\left[\left(1-\frac{4}{3}\beta^{2}\right)x^{2}-y^{2}-\frac{4}{3}\beta x\right]+\frac{4}{3}\beta\beta_{{\rm dm}}\left[1-(1-4\beta^{2})x^{2}-y^{2}+4\beta x-z^2\right]+\frac{2\sqrt{6}}{3}\beta\mu\,y^{2} \,. \label{eq: H'H-1-1-1}
\end{align}
Continuing with the Klein-Gordon equation:
\begin{equation}
\phi''+\left(3+\frac{H'}{H}\right)\phi'+\frac{1}{2}\frac{f_{1,\phi}}{f_{1}}\phi'{}^{2}=-\frac{1}{(1-4\beta^{2})\,f_{1}H^{2}}\left[V_{,\phi}-\frac{3f_{1,\phi}}{\kappa}H^{2}\left(2+\frac{H'}{H}\right)+\frac{f_{2,\phi}}{2f_{2}}\rho_{{\rm dm}}\right]\,,\label{eq:KG_BG-1}
\end{equation}
which can be written as
\begin{align}
x'+\left(3+\frac{H'}{H}\right)x+2\beta x^{2} & =\frac{3\mu y^{2}}{(1-4\beta^{2})\sqrt{6}\,}+\frac{2}{1-4\beta^{2}\,}\beta\left(2+\frac{H'}{H}\right)+\beta_{{\rm dm}}\frac{\tilde{\Omega}_{{\rm dm}}}{1-4\beta^{2}\,}\,.\label{eq:KG_BG-1-1-1}
\end{align}
Then, from the definition of $y$ and $z$, we also have
\begin{align}
\frac{y'}{y} & =-\frac{H'}{H}+\frac{1}{2}\frac{V'}{V}-\frac{1}{2}\frac{f_{1}'}{f_{1}} \nonumber\\
 & =-\frac{H'}{H}-\frac{1}{2}\mu\sqrt{6}x-2\beta x\, , \\
 \frac{z'}{z}&=-\frac{3}{2}-\frac{H'}{H}-2\beta x \,.
 \end{align}
So the final system of autonomous equations is:
\begin{align}
x' & =-\left(3+\frac{H'}{H}\right)x-2\beta x^{2}+\frac{\sqrt{6}\mu y^{2}}{2(1-4\beta^{2})\,}+\frac{2}{1-4\beta^{2}\,}\beta\left(2+\frac{H'}{H}\right)+\beta_{{\rm dm}}\frac{\tilde{\Omega}_{{\rm dm}}}{1-4\beta^{2}\,}\,,\label{eq:system1}\\
y' & =y\left(-\frac{H'}{H}-\frac{1}{2}\mu\sqrt{6}x-2\beta x\right)  \,,  \label{eq:system2}\\
z' & =z\left( -\frac{3}{2}-\frac{H'}{H}-2\beta x\right)\,.
\label{eq:system3}
\end{align}
For $\beta=0$ we get
\begin{align}
x' & =-\left(3+\frac{H'}{H}\right)x+\frac{\sqrt{6}\mu y^{2}}{2\,}+\beta_{{\rm dm}}(1-x^{2}-y^{2}-z^{2})\,,\\
y' & =y\left(-\frac{H'}{H}-\frac{\sqrt{6}}{2}\mu x\right) \,,  \\
z' & = z\left( -\frac{3}{2}-\frac{H'}{H}\right)  \,.
\end{align}
which coincides with the usual equations of CDE, with
\begin{align}
\frac{H'}{H} & =-\frac{3}{2}(1+x^{2}-y^{2})
\end{align}
Now we find the critical points of the system Eqs. \eqref{eq:system1}, \eqref{eq:system2} for $y=0$ and $z=0$:
\begin{align}
x' & =-\left(3+\frac{H'}{H}\right)x-2\beta x^{2}+\frac{2\beta}{1-4\beta^{2}\,}\left(2+\frac{H'}{H}\right)+\beta_{{\rm dm}}\frac{1-(1-4\beta^{2})x^{2}+4\beta x}{1-4\beta^{2}\,}\,,\\
y' & =0  \,, \\
z' & =0   \, ,
\end{align}
with
\begin{equation}
-\frac{2}{3}\frac{H'}{H}=1+\frac{4}{3}\beta^{2}+(1-4\beta^{2})\left[\left(1-\frac{4}{3}\beta^{2}\right)x^{2}-\frac{4}{3}\beta x\right]+\frac{4}{3}\beta\beta_{{\rm dm}}\left[1-(1-4\beta^{2})x^{2}+4\beta x\right] \, .
\end{equation}
Then we can simplify the $x'$ equation and obtain
\begin{equation}
x'=\frac{1}{2}\left(\left(4\beta^{2}-1\right)x^{2}+4\beta x+1\right)\left(2\beta+\beta_{\text{dm}}(4\beta x+2)+\left(4\beta^{2}-3\right)x\right) \, .
\end{equation}
The $y=0$ critical points  are then given by solving $x'=0$:
\begin{align}
x_{\pm} & =\frac{1}{\pm1-2\beta}  \, , \\ 
x_{i} & =\frac{2\left(\beta+\beta_{\text{dm}}\right)}{3-4\beta(\beta+\beta_{\text{dm}})}  \, .
\end{align}
Since these critical points exist for $y=0$, they are independent of the particular choice of the  potential.
On the two points $x_\pm$, $\tilde\Omega_{\rm dm}=0$, so they cannot be valid early-time solutions. Instead, the last point yields a finite amount of dark matter and early dark energy, and should be taken as initial condition, as we have shown in the previous appendix.

A detailed analysis of the full phase space and of the stability of the critical points is left to future work.

\section{Details on the Fisher Matrix Projection} \label{app: Projection}

In this appendix, we present details on the Fisher matrix projection. More specifically, we project the model-independent constraints on the parameters $\left\{E_P, h_d, d\right\}$, obtained using a methodology similar to that in Refs. \cite{EP, Zheng:2025oiq} (with each parameter left free to vary at each redshift bin), onto a set of constant parameters: 
$\left\{\beta, \beta_{\rm dm}, \Omega_{\rm m0}\right\}$. When all three parameters are varied, we obtain constraints on the doubly coupled dark energy model, while in the case $\beta=0$, the projection corresponds to the coupled dark energy model. This can be done by transforming the original Fisher matrix, $\mathcal{F}_{\textrm{old}}$, which contains only the information on $\left\{E_P, h_d, d\right\}$ at different redshift bins, into a new $3\times3$ Fisher matrix, $\mathcal{F}_{\textrm{new}}$, from whose inverse we obtain the marginalised errors on $\left\{\beta, \beta_{\rm dm}, \Omega_{\rm m0}\right\}$, as follows:
\begin{equation}
    \mathcal{F}_{\textrm{new}}=J^{T}\mathcal{F}_{\textrm{old}}J \,,
\end{equation}
with $J$ the Jacobian matrix and $J^T$ its transpose. For simplicity, hereafter we define that, in the i-th redshift bin, $E_P(N_i)\equiv E_{P,i}$, $h_d(N_i)\equiv h_{d,i}$ and $d(N_i)\equiv d_i$. The $E_P(N)$ expressions for our specific models are given by Eqs. \eqref{eq:epfin} and \eqref{eq: Ep_xy_DCDE}, while those for $d(N)$ and $h_d(N)$ are given by
\begin{align}
    d(N) &= \frac{L_A(N)}{L_{A_r}(N)}
    = \frac{\int_{N}^{0}\frac{e^{-N'}\,\mathrm{d}N'}{E(N')}}
    {\int_{N}^{0}\frac{e^{-N'}\,\mathrm{d}N'}{E_r(N')}}  \,, \\
    h_d(N) &= h(N)d(N)
    = \frac{E(N)}{E_r(N)}\frac{L_A(N)}{L_{A_r}(N)}
    = \frac{E(N)\int_{N}^{0}\frac{e^{-N'}\,\mathrm{d}N'}{E(N')}}
    {E_r(N)\int_{N}^{0}\frac{e^{-N'}\,\mathrm{d}N'}{E_r(N')}} \,,
\end{align}
where $E_r$ is the expansion rate for the reference cosmology, i.e. the $\Lambda$CDM model. The expansion rate $E$, on the other hand, depend on $\beta$ and $\beta_{\rm dm}$ through $x$ and $y$ in Eqs. \eqref{eq: x_QQdm} and \eqref{eq: V_QQdm}, with the exact solution given by Eq. \eqref{eq: E_exact}. Moreover, in the CDE case, $E(N)$ simply reduces to
\begin{align} \label{eq: APPC_E_CDE}
    E(N)= e^{-3/2 \int_{0}^{N} (x(N')^2-y(N')^2+1)dN'}  \, ,
\end{align}
with the $x$ and $y$ given by Eqs. \eqref{eq:x} and \eqref{eq:y}. This expansion is well justified, as the relative difference from $E_r(N)$ remains below $0.4\%$ over the low-redshift range considered in our forecast, as shown in Table~\ref{tab:E_comparison}.
\begin{table*}
\centering
\begin{tabular}{cccc}
\hline\hline
$z$ & $E_r$ & $E$ & $\left|E-E_r\right|/E_r$ [\%] \\
\hline
0.10 & 1.050840 & 1.051463 & 0.0593 \\
0.30 & 1.173480 & 1.174974 & 0.1274 \\
0.50 & 1.322167 & 1.324351 & 0.1652 \\
0.70 & 1.494187 & 1.497100 & 0.1949 \\
0.90 & 1.686886 & 1.690686 & 0.2253 \\
1.10 & 1.897950 & 1.902861 & 0.2588 \\
1.30 & 2.125466 & 2.131747 & 0.2955 \\
1.50 & 2.367884 & 2.375817 & 0.3352 \\
\hline\hline
\end{tabular}
\caption{Comparison of the reference expansion rate $E_r$ and the expanded expression for $E$ at different redshifts. The last column shows the relative difference between the two expressions.}
\label{tab:E_comparison}
\end{table*}

The Jacobian elements $J^{(i)}$ within each redshift bin $\eta_i$ are given by: 
\begin{equation}
J^{(i)}=
\begin{pmatrix}
\displaystyle \frac{\partial h_{d,i}}{\partial \beta} &
\displaystyle \frac{\partial h_{d,i}}{\partial \beta_{\rm dm}} &
\displaystyle \frac{\partial h_{d,i}}{\partial \Omega_{\rm m0}}

\\[1.5ex]
\displaystyle \frac{\partial d_i}{\partial \beta} &
\displaystyle \frac{\partial d_i}{\partial \beta_{\rm dm}} &
\displaystyle \frac{\partial d_i}{\partial \Omega_{\rm m0}}
\\[1.5ex]
\displaystyle \frac{\partial E_{P,i}}{\partial \beta} &
\displaystyle \frac{\partial E_{P,i}}{\partial \beta_{\rm dm}} &
\displaystyle \frac{\partial E_{P,i}}{\partial \Omega_{\rm m0}}
\end{pmatrix} \,,
\end{equation}
with each entry evaluated at the fiducial values, e.g: 
\begin{equation}
\left.\frac{\partial E_{P}(N_i)}{\partial \beta}\right|_{\rm fid}
=
\left.
\frac{\partial}{\partial \beta}
\Big[
E_P(N_i;\beta,\beta_{\rm dm},\Omega_{\rm m0})
\Big]
\right|_{
N=N_i;\,
\beta=\beta_{\rm fid},\,
\beta_{\rm dm}=\beta_{\rm dm, fid},\,
\Omega_{\rm m0}=\Omega_{\rm m0, fid}
}.
\end{equation}
In this paper we set: $\beta_{\rm fid}=0.1$, $\beta_{\rm dm, fid}=0.2$, and $\Omega_{\rm m0,\rm fid}=0.315$.

For the full Jacobian $J$ including all the number of $N$ redshift bins, the blocks are stacked as: 
\begin{equation}
J=
\begin{pmatrix}
J^{(1)}\\
J^{(2)}\\
\vdots\\
J^{(N)}
\end{pmatrix} \,.
\end{equation}

\section{Estimates of the relative error on the luminosity distance}
\label{app:sn}

We give now a rough estimate of the current uncertainty on the luminosity distance $d$ that we use as prior in the main text. For $N$ independent supernovae with identical Gaussian apparent-magnitude uncertainty $\delta m$ and perfectly known absolute magnitude, the relative error on $d$ is
\begin{equation}
\frac{\sigma_{d}}{d}=\frac{\ln 10}{5}\frac{\delta m}{\sqrt{N}}.
\end{equation}
Using the current Pantheon+ catalog \cite{Brout:2022vxf}, we show the resulting uncertainty on $d$ in Table \ref{tab:sn}

\begin{table}
\centering

\label{tab:pantheon_distance_errors}
\begin{tabular}{ccccc}
\hline\hline
Bin & Redshift range & $N$ &
$\delta d/d$ ($\delta m=0.10$) &
$\delta d/d$ ($\delta m=0.15$) \\
\hline
1 & $0.0-0.1$ & 583 & $0.19\%$ & $0.28\%$ \\
2 & $0.1-0.2$ & 207 & $0.32\%$ & $0.48\%$ \\
3 & $0.2-0.3$ & 259 & $0.29\%$ & $0.43\%$ \\
4 & $0.3-0.4$ & 186 & $0.34\%$ & $0.51\%$ \\
5 & $0.4-0.5$ &  98 & $0.47\%$ & $0.70\%$ \\
6 & $0.5-0.6$ &  81 & $0.51\%$ & $0.77\%$ \\
7 & $0.6-0.7$ &  54 & $0.63\%$ & $0.94\%$ \\
8 & $0.7-0.8$ &  45 & $0.69\%$ & $1.03\%$ \\
\hline\hline
\end{tabular}
\caption{Pantheon+ supernova luminosity-distance uncertainties.\label{tab:sn}}
\end{table}

These results are just a rough estimate, since we  assumed a constant $\delta m$ and ignored  correlated systematics, peculiar velocities, and intrinsic dispersion. However, they also show that within the SKA redshift range, our prior of 1\% is  justified, and will  be soon improved by further data.

\bibliographystyle{ieeetr} 
\addcontentsline{toc}{section}{References}
\clearpage
\bibliography{references}
\end{document}